\documentclass[sigconf]{acmart}

\setcopyright{none}
\renewcommand\footnotetextcopyrightpermission[1]{}
\usepackage{graphicx} % Required for inserting images
\usepackage{booktabs}
\usepackage{multirow}
\usepackage{ulem}
\usepackage{colortbl}
\usepackage{subcaption}
\usepackage[ruled,vlined,linesnumbered]{algorithm2e}
\usepackage{color}
\usepackage{listings}

\definecolor{GrayCodeBlock}{RGB}{241,241,241}
\definecolor{BlackText}{RGB}{48,10,2}
\definecolor{RedTypename}{RGB}{182,86,17}
\definecolor{GreenString}{RGB}{96,172,57}
\definecolor{PurpleKeyword}{RGB}{165,5,225}
\definecolor{GrayComment}{RGB}{92, 91, 93}
\definecolor{GoldDocumentation}{RGB}{180,165,45}
\lstdefinelanguage{rust}
{
    columns=fullflexible,
    keepspaces=true,
    frame=single,
    framesep=0pt,
    framerule=0pt,
    framexleftmargin=4pt,
    framexrightmargin=4pt,
    framextopmargin=5pt,
    framexbottommargin=3pt,
    xleftmargin=4pt,
    xrightmargin=4pt,
    backgroundcolor=\color{GrayCodeBlock},
    basicstyle=\footnotesize\ttfamily\color{BlackText},
    morekeywords=[1]{unsafe},
    morekeywords=[2]{int, float},
    morekeywords=[3]{struct, fn, let, return, if, else, while, impl, pub},
    keywordstyle=[1]\color{red},
    keywordstyle=[2]\color{blue},
    keywordstyle=[3]\color{PurpleKeyword},
    ndkeywords={
        bool,u8,u16,u32,u64,u128,i8,i16,i32,i64,i128,char,str,
        Self,Option,Some,None,Result,Ok,Err,String,Box,Vec,Rc,Arc,Cell,RefCell,HashMap,BTreeMap,
        macro_rules
    },
    ndkeywordstyle=\color{RedTypename},
    comment=[l][\color{GrayComment}\slshape]{//},
    morecomment=[s][\color{GrayComment}\slshape]{/*}{*/},
    morecomment=[l][\color{GoldDocumentation}\slshape]{///},
    morecomment=[s][\color{GoldDocumentation}\slshape]{/*!}{*/},
    morecomment=[l][\color{GoldDocumentation}\slshape]{//!},
    morecomment=[s][\color{RedTypename}]{\#![}{]},
    morecomment=[s][\color{RedTypename}]{\#[}{]},
    stringstyle=\color{GreenString},
    string=[b]"
}

\acmConference[]{arXiv}{2024}{}
\title{Characterizing Unsafe Code Encapsulation In Real-world Rust Systems}
\author{Zihao Rao}
\affiliation{%
  \institution{School of Computer Science,\\ Fudan University}
  \country{China}
}

\author{Yiran Yang}
\affiliation{%
  \institution{School of Computer Science,\\ Fudan University}
  \country{China}
}

\author{Hui Xu}
\authornote{Corresponding author.}
\affiliation{%
  \institution{School of Computer Science,\\ Fudan University}
 \country{China}
}

\begin{abstract}
Interior unsafe is an essential design paradigm advocated by the Rust community in system software development. However, there is little official guidance or few best practices regarding how to encapsulate unsafe code and achieve interior unsafe. The problem is critical because the Rust compiler is incapable of verifying the soundness of a safe function containing unsafe code. Falsely declaring an interior unsafe function as safe may undermine the fundamental memory-safety guarantee of Rust. To address this issue, this paper studies how interior unsafe is achieved in practice, aiming to identify best practices to guide Rust code design concerning unsafe code encapsulation. Specifically, we propose a novel unsafety isolation graph to model the essential usage and encapsulation of unsafe code. Based on the graph, we further propose four major isolation types and nine structural patterns to split a graph into several small self-contained subgraphs. These subgraphs can serve as useful audit units for examining the soundness of unsafe code encapsulation. We applied our approach to four real-world Rust projects. The experimental results demonstrate that our method is effective in characterizing their encapsulation code. Additionally, we identified two common issues in these projects that could complicate soundness verification or incur unsoundness issues.

\end{abstract}
\begin{document}

\maketitle

\section{Introduction}
% Paragraph 1: 1) Providing safe API to users is an important goal for Rust developers. 2) How developers encapsulate unsafe code as safe API in real-world projects is not studied. 2) The problem is important because the safety of Rust depends on the declaration correctness of its interior unsafe API. 
%The Rust programming language has gained substantial popularity among systems developers, due to its advantages in memory safety and efficiency. Many system software, such as the operating system Asterinas~\cite{asterinas_github} and Redox~\cite{redox_website}, have adopted Rust as their primary programming language. However, 

The Rust programming language has two parts: safe Rust and unsafe Rust~\cite{safeandunsafe}. Leveraging novel designs like ownership and borrow checking, the Rust compiler can ensure memory safety as long as developers do not use unsafe code~\cite{jung2019stacked,evans2020rust}. However, avoiding the use of unsafe code is generally impossible for system software development, which requires low-level operations involving raw pointer arithmetic and dereferencing. Consequently, interior unsafe has become an essential design paradigm advocated by the Rust community~\cite{mccormack2024against,qin2024understanding}. This paradigm suggests that developers should strive to encapsulate unsafe code within safe functions or methods whenever possible. Meanwhile, since unsafe code can lead to undefined behaviors, developers must ensure the soundness of their encapsulated interior unsafe functions themselves, as the compiler is unable to verify this. Unsoundness issues in interior unsafe functions are common and represent a typical type of bug unique to Rust~\cite{xu2021memory,mccormack2024against}.

Despite the importance of interior unsafe, there is little guidance from the Rust community on how to achieve sound encapsulation. Prior to this work, several empirical studies (\textit{e.g.,} ~\cite{evans2020rust,astrauskas2020programmers}) have focused on how developers employ unsafe code, rather than how to encapsulate it or achieve interior unsafe. Additionally, there are investigations (\textit{e.g.,} ~\cite{matsushita2022rusthornbelt,bae2021rudra,astrauskas2022prusti}) into soundness verification or detecting unsoundness bugs. However, these studies provide very limited insights for developers in designing their code. To the best of our knowledge, there are no systematic studies on how to design interior unsafe code and how interior unsafe is implemented in real-world projects. 

In this paper, we aim to study how interior unsafe is achieved in practice and identify best practices for interior unsafe code design. The problem is challenging because Rust has a complicated type system with structs and traits. Developers may have to consider multiple functions to examine the soundness of an interior unsafe function. For example, an interior unsafe method of a struct may depend on its constructor to be safe, and the struct may have several different constructors and interior unsafe methods. Such dependencies can create a strong coupling relationship among multiple interior unsafe code. Additionally, some interior unsafe code may rely on global states or external functions to avoid undefined behaviors, which further complicates the soundness verification.

To address this challenge, we propose a novel \textit{unsafety isolation graph} to track the propagation of unsafe code and identify functions responsible for isolating unsafety. The graph captures the essential relationships among constructors, methods, and functions regarding unsafe code encapsulation. Given that the graph can become very large and complex in some projects, we further propose four isolation types and nine structural patterns to divide the graph into smaller, manageable subgraphs. Each subgraph acts as a self-contained audit unit for unsafety encapsulation and verification. We have developed an audit formula for each pattern to guide developers in examining the soundness of their interior unsafe code. Our approach is systematic and comprehensive, considering all possible scenarios of dependencies and isolation responsibilities involving structs and traits.
%For instance, direct isolation involves only one hop on the unsafety isolation graph for verification, while full isolation involves two hops. Half isolation contains two separate units for verification, with only one unit being self-contained if considering the safety descriptions. Open isolation has two or three separate units, and it also has only one unit being self-contained  unit for verification if leveraging safety descriptions. Our categorization approach is comprehensive, as it considers all possible situations of dependencies and isolation responsibilities involving structs and traits.

We applied our approach to four renowned operating system projects written in Rust. Experiment results show that our approach is very useful for auditing interior unsafe code. In particular, we find two common issues in unsafe code encapsulation. Firstly, developers may forget to specify the safety requirements for unsafe constructors, complicating the verification of the soundness of dynamic methods associated with these constructors. Secondly, developers may ignore the literal constructor of a struct, which can bypass the safety requirements of its unsafe constructor, leading to unsound interior unsafe code.

In short, our paper makes the following contributions.

\begin{itemize}
    \item We propose a novel approach for auditing unsafe code encapsulation in Rust projects. Our method decomposes complex interior unsafe code into nine structural patterns and provides an audit formula for each pattern. This approach offers valuable guidance for developers in designing robust interior unsafe code.
    \item We conducted experiments on four real-world Rust projects and identified two common issues that may lead to unsound interior unsafe code design. We believe our findings and suggestions will be valuable to the Rust community. 
\end{itemize}

\section{Preliminary}
In this section, we present the background of unsafe Rust and interior unsafe code.

\subsection{Overview of Rust}
Rust has two parts: safe Rust and unsafe Rust. 

\subsubsection{Safe Rust}
Safe Rust guarantees developers that if a program without unsafe code compiles, it will not exhibit undefined behavior. Rust achieves this through several innovative features, including ownership and lifetimes. In safe Rust, each value has an owner, and ownership can be borrowed either mutably or immutably. The Rust compiler tracks ownership and enforces the rule of exclusive mutability: if a value has a mutable reference at a given point, it cannot have other mutable or immutable aliases at that same point. However, safe Rust has its limitations. For instance, implementing a doubly linked list is challenging because each node needs to have two aliases, and the node itself must be mutable.

\subsubsection{Unsafe Rust}
To enhance usability, Rust defines five types of unsafe code: raw pointer dereference, calling an unsafe function or method, accessing a static mutable variable, accessing fields of unions, and implementing an unsafe trait. Because the Rust compiler cannot guarantee memory safety for unsafe code, such code must be used within an unsafe block (\textit{i.e.,} annotated with an \texttt{unsafe} marker), indicating that developers should be responsible for the risks incurred by unsafe code. Otherwise, the Rust compiler will reject the code. Unsafe Rust complements safe Rust by providing necessary low-level memory control and flexible interoperability for many system software development scenarios.

\subsection{Interior Unsafe and Challenges}
Since unsafe code is sometimes unavoidable in Rust programming, it is crucial to prevent the uncontrolled propagation of unsafety risks. Interior unsafe is an essential design paradigm to mitigate these risks. This paradigm involves a safe function or method containing unsafe code, where all potential undefined behaviors caused by the unsafe code are properly managed within the function itself~\cite{qin2020understanding,qin2024understanding}. For example, Listing~\ref{lst:unsafecase1} demonstrates an interior unsafe function \texttt{foo()} that contains a call to the unsafe function \texttt{doUnsafe(x)}. The unsafe code in  \texttt{foo()} is only executed if \texttt{x} passes a validity check. Developers ensure that regardless of the argument passed to \texttt{foo()}, the function does not produce any undefined behaviors. Thus, interior unsafe is an effective strategy to prevent the propagation of unsafety risks and is widely used in Rust libraries.

\begin{lstlisting}[language=rust, caption={A simple interior unsafe function.}, label={lst:unsafecase1}]
fn foo(x: MyType) { 
    if (IsValid(x)) { 
        unsafe { doUnsafe(x); }
    }
}
\end{lstlisting}

\begin{lstlisting}[language=rust, caption={Interior unsafe Rust code with struct.}, label={lst:unsafecase2}]
struct MyStruct<T> {
    x: *T,
    n: i32,
}
impl struct MyStruct<T> {
    fn new(x: *T) -> Self { ... }
    unsafe fn deref(&self) -> T { ... }
    fn foo(&self) { unsafe{...} }
    fn bar(&self) { unsafe{...} }
}
\end{lstlisting}

Currently, the soundness check of interior unsafe code primarily relies on developers. While this task may be manageable for small-scale projects with only a few unsafe code snippets, it becomes challenging in real-world Rust projects as the number of unsafe code snippets grows. Real-world Rust projects often involve complex structs and traits, requiring developers to consider all combinations of member functions to ensure soundness. Listing presents a struct \texttt{MyStruct} with a safe constructor \texttt{new}, an unsafe member function \texttt{deref()}, and two safe methods \texttt{foo()} and \texttt{bar()}. Developers must not only verify that \texttt{new()}, \texttt{foo()}, and \texttt{bar()} are individually sound but also ensure that their combinations in different orders are sound. Additionally, the safety requirements for unsafe code may not be explicit or well-documented, and validity checks might involve complex logic, such as external functions or global variables. These challenges can lead to unsoundness in interior unsafe code.

\begin{figure*}[h]
    \centering
    \includegraphics[width=0.8\linewidth]{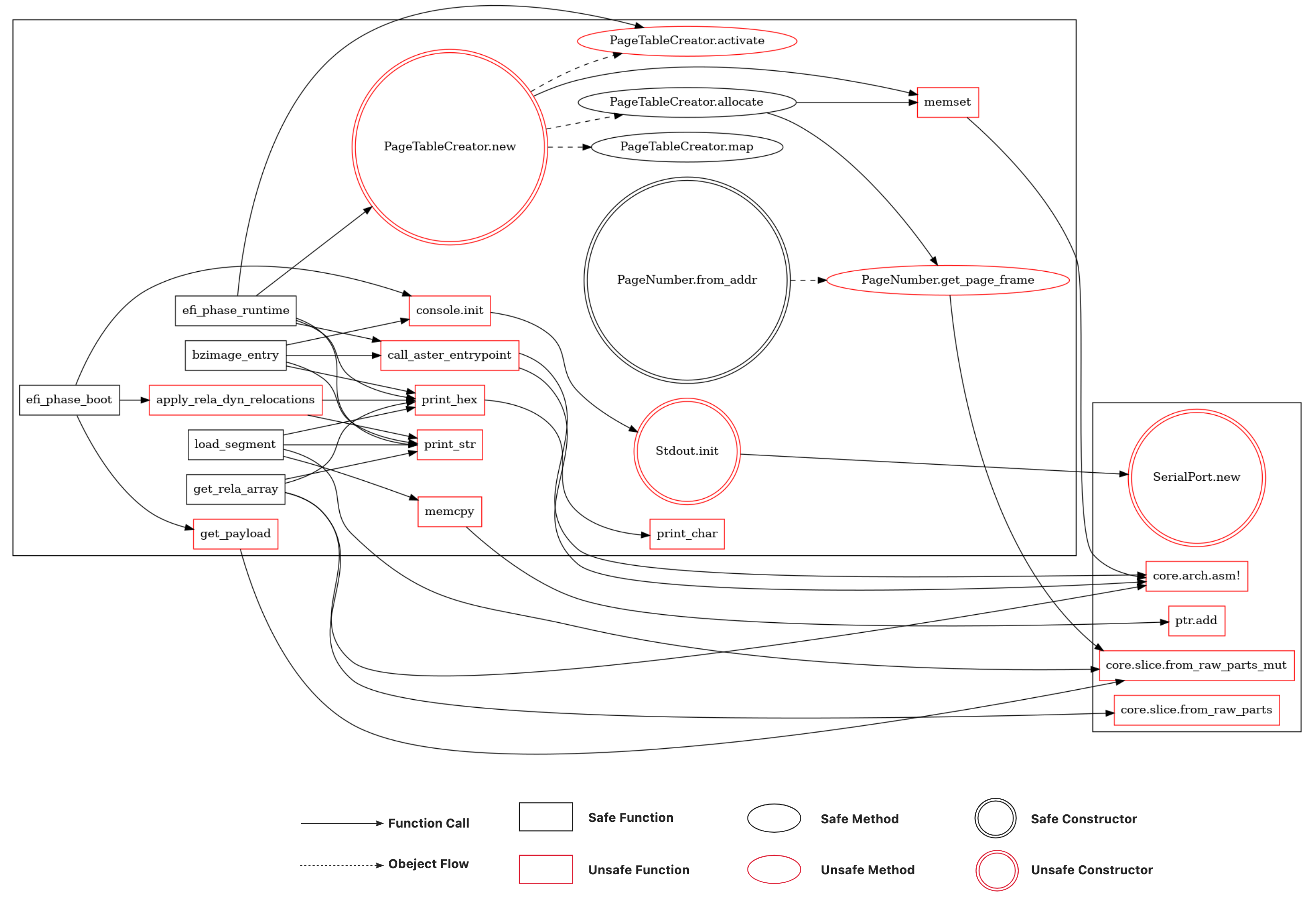}
    \caption{A sample UPG for the Asterinas project.}
    \label{fig:UPG_overview}
\end{figure*}

\section{Characterizing Unsafety Isolation}\label{sec:method}
\subsection{Problem Setting}
In this paper, we aim to study how interior unsafe code is encapsulated and to identify best practices for its safe usage. Safety encapsulation is relatively straightforward for interior unsafe functions that are not part of any structs or traits, as they only depend on the function’s internal logic. However, the complexity increases when dealing with structs and traits. In this section, we clarify several key terminologies and mechanisms related to Rust.

\subsubsection{Structs and Traits}
A struct in Rust is a data structure that contains data fields, constructors, dynamic methods, and static methods. Each struct has a default literal constructor that can create an instance by directly specifying the value of each data field. Dynamic methods are different from static methods in that their first parameter is \texttt{self}~\cite{rustfunctions}. Dynamic methods differ from static methods in that their first parameter is \texttt{self}. A dynamic method can only be used after an instance of the struct has been created and is called using the syntax \texttt{object.foo()}, where \texttt{\&self} points to the instance. In contrast, static methods do not require an instance to be created and are called directly using the syntax \texttt{MyStruct::zoo()}. A trait is a collection of methods that can be implemented by structs or shared among them. Traits define common behaviors that structs can adopt, enabling polymorphism and code reuse in Rust.

\subsubsection{Unsafe Marker}
There are two primary ways to use unsafe code in Rust. One declares the entire function as unsafe, allowing the use of unsafe code anywhere within it. This makes calling the function itself unsafe, indicating that the caller must ensure safety when using this function. The other way only marks a specific block within a function as unsafe, while the function itself can be declared as either safe or unsafe. The Rust compiler can verify that unsafe code is used only within the designated unsafe scope. However, it cannot ensure that a function declared as safe but containing unsafe code within such a scope is truly safe in all scenarios.

\subsubsection{Namespace and Visibility}
In Rust, a crate is the smallest compilation unit used to build targets, such as executables and libraries. A crate can contain multiple modules, each representing a different namespace. Modules can be further divided into sub-modules. By default, all program components (\textit{e.g.,} functions and global variables) within a module are private, meaning they are accessible only within that module. However, developers can make these components public by using the \texttt{pub} keyword, allowing them to be accessed from outside the module.

\subsection{Modeling Unsafety with Graphs}\label{modeling}
This section introduces a novel approach for modeling essential program components to facilitate unsafe code encapsulation. Traditional call graphs fall short of our requirements because they do not distinguish between safe functions, unsafe functions, and interior unsafe functions. Consequently, they do not provide the necessary information to identify suitable functions for encapsulating unsafe code based on call graphs alone. To address this limitation and enhance our research on interior unsafe code, we propose a novel graph called the Unsafety Propagation Graph (UPG).

\begin{definition}[Unsafety Propagation Graph]
A UPG $\mathcal{G(S,U,C,E)}$ is a directed graph that tracks the propagation of unsafe code through function calls and objects in a Rust program.

\begin{align*}
\mathcal{S} & = S_f\cup S_m \\
\mathcal{U} & = U_f\cup U_m \\
\mathcal{C} & = C_s\cup C_u \cup C_l \\
\mathcal{E} & = E_{obj}\cup E_{call}
\end{align*}

where

\begin{itemize}
    \item $S_f$ is a set of safe functions and safe static methods.
    \item $S_m$ is a set of safe dynamic methods.
    \item $U_f$ is a set of unsafe functions and unsafe static methods.
    \item $U_m$ is a set of unsafe dynamic methods.
    \item $C_s$ is a set of safe constructors.
    \item $C_u$ is a set of unsafe constructors.
    \item $C_l$ is a set of literal constructors.
    \item $E_{obj}$ represents object flows between constructors and dynamic methods.
    \item $E_{call}$ represents function calls with an unsafe callee.
\end{itemize}
\end{definition}

\begin{algorithm}
\small
    \caption{Unsafety Property Graph Generation}
    \label{upg_generation}
    \SetAlgoLined
    \KwIn{$crate$: The source code of a Rust crate}
    \KwOut{$UPG$: Unsafety Property Graph}
    $UPG$ $\leftarrow \emptyset$ \;
    $Nodes$ $\leftarrow$ FilterByUnsafeMarker($crate$)\;
    $UPG$.AddNode($Nodes$)\;
    \ForEach{$node$ in $Nodes$}
    {
        \If{$node$.Type == DynMethod}{
            $Constructors$ = FindConstructors($node$)\;
            \ForEach{$cons$ in $Constructors$}{
                $UPG$.AddObjFlow($cons$, $node$)\; 
            }
        }
        CallSites $\gets$ FindAllUnsafeCallSites($node$)\;
        \ForEach{$site$ in $CallSites$}{
            $Callees$ $\gets$ $\emptyset$\;
            \If{$site$.calleeType == DynMethod}{
                $insType$ $\gets$ GetInstanceType($site.callee$)\;
                \If{$insType$ == Generic}{ 
                    $Callees$ $\gets$   FindAllImpls($site.callee$)}
                \Else {
                    $Callees$ $\gets$ $site.callee$ }
                }  
                \ElseIf{$site$.calleeType == FnParam}{
                    $fnSign$ $\gets$   GetFnSign($site.callee$)\;
                    $Callees$ $\gets$   FindAllFns($fnSign$)
                }
                \Else {
                    $Callees$ $\gets$ $site.callee$ }
                \ForEach{$callee$ in $Calless$}{
                    $UPG$.AddEdge($site.caller$, $callee$)\;
                }
            }
        }
    
\end{algorithm}

Note that in the UPG, we are interested in two kinds of function calls: a safe caller and an unsafe callee, or an unsafe caller with an unsafe callee. Additionally, we treat static methods in the same way as general functions because they can be invoked directly without creating an instance of the struct, and hence do not affect other methods. However, dynamic methods cannot be invoked without a struct instance. Therefore, we should also consider the relationship between constructors and dynamic methods, denoted as $E_{obj}$. Since the objective of a UPG is to track unsafety propagation, we do not need to consider unrelated function calls among safe functions. 

Figure~\ref{fig:UPG_overview} illustrates a sample UPG for the real-world Rust project Asterinas. To better visualize unsafe code propagation paths, we use different colors and shapes to represent various functions and relations. Specifically, red-colored nodes indicate unsafe functions. Additionally, we distinguish between functions, methods, and constructors using rectangle, oval, and circle shapes, respectively. Solid lines represent function calls, while dashed lines indicate object flows between dynamic methods and their constructors.

Algorithm~\ref{upg_generation} demonstrates the process of UPG extraction. It begins by identifying all functions and methods with unsafe markers and adds them to the UPG as nodes. For each dynamic method, the algorithm locates its constructors and adds object-flow edges between them. Next, it searches for all unsafe call sites within the nodes and adds corresponding call edges. If a callee is an unsafe dynamic method with a generic instance type, it implies the potential for multiple method implementations. In such cases, the method must belong to a trait that bounds the generic parameter, and the algorithm searches for all implementations of the trait method, adding corresponding edges. Similarly, if an unsafe callee originates from a function argument, the algorithm considers all possible functions with the same signature in the UPG.

While a UPG can visualize unsafety propagation, it may contain redundant elements concerning safe encapsulation. To address this, we proceed to remove all unsafe nodes that have no callers in the UPG, resulting in the creation of an Unsafety Isolation Graph (UIG).

\begin{definition}[Unsafety Isolation Graph] An UIG is an unsafety propagation graph that has no $U_f$ or $U_m$ without callers.
\end{definition}

Within each UIG, every unsafe node carries specific safety requirements that must be met ~\cite{cui2024unsafe}. For instance, the unsafe method \texttt{vec::set\_len} from the Rust standard library specifies that before calling, the argument \texttt{new\_len} must be less than or equal to \texttt{self.capacity}, and the elements from index \texttt{old\_len} to \texttt{new\_len} must be initialized. We use the term required safety property set (RS) to denote such requirements. Correspondingly, each safe caller on a UIG must ensure that it satisfies its callee's RS to avoid undefined behaviors. We refer to the safety ensured by the caller as the verified safety property set (VS). For instance, a safe caller to \texttt{vec::set\_len} must assert that \textit{new\_len} is less than \textit{vec.capacity} and guarantee that the extended content is well-initialized.

On a UIG, we do not consider safe callers with safe callees because the responsibility for ensuring soundness lies with the safe callee regardless of the caller. This approach eliminates the need to account for various call sequences. Specifically, if a safe method contains unsafe code, we do not have to consider different combinations of their methods.

\subsection{Structural Patterns of Unsafety Isolation}

While an unsafety isolation graph (UIG) can capture the essential components for unsafe code encapsulation, it may become too large (\textit{e.g.,} Figure~\ref{fig:UPG_overview}) for effective auditing. In this section, our objective is to propose a method for splitting a UIG into multiple self-contained subgraphs for auditing purposes. 

In our approach, each audit unit consists of a pair comprising a safe caller with an unsafe callee, along with their constructors if they are dynamic methods. We categorize all audit units into four major folds based on their structural patterns and verification considerations.

\subsubsection{Direct Isolation}
This is a straightforward pattern that only the safe function itself is responsible for encapsulating the unsafe code. This type includes two subtypes.
\begin{figure}[H]
    \centering
    \begin{subfigure}[b]{0.22\textwidth}   
        \includegraphics[width=\textwidth]{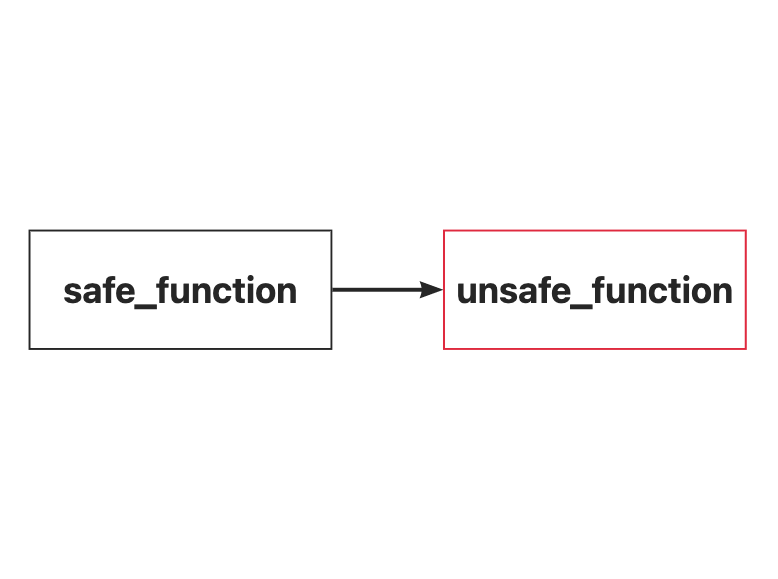}
        \label{fig:sf_uf}
        \vspace{-0.6cm}
        \caption{Type I} 
    \end{subfigure}
    \hfill
    \begin{subfigure}[b]{0.22\textwidth}
        \includegraphics[width=\textwidth]{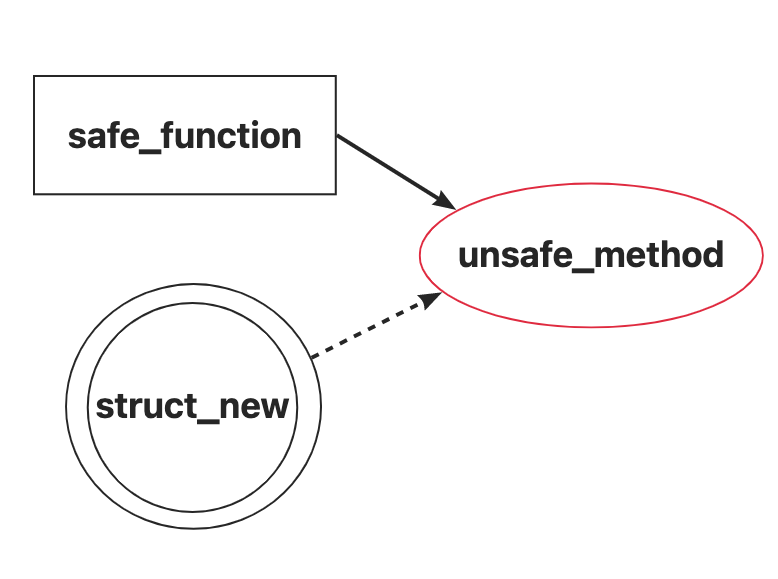}
        \label{fig:sf_um}
        \vspace{-0.6cm}
        \caption{Type II} 
    \end{subfigure}
    \caption{Direct Isolation}
\end{figure}

\begin{itemize}
    \item $s_f\to u_f||c_u$: A safe function $s_f$ calls either an unsafe function $u_f$ or an unsafe constructor $c_u$. It is the responsibility of $s_f$ to ensure that all the safety requirements of $u_f$ can be satisfied when reaching $u_f$. It's important to note that since constructors are also static functions, $u_f$ can also represent an unsafe constructor $c_u$. Audit formula: 
    \[RS_{u_f} \subseteq VS_{s_f} \]
    
    \item $s_f||c_s\to u_m(c_s)$: A safe function $s_f$ calls an unsafe method $u_m$, and the constructor $c_s$ associated with $u_m$ is safe. Since $u_m$ has specific safety requirements, $s_f$ must ensure that all these requirements are satisfied. Audit formula: 
    \[RS_{u_m} \subseteq VS_{s_f} \]
\end{itemize}

\begin{figure}[H]
    \centering
    \begin{subfigure}[b]{0.22\textwidth}
        \includegraphics[width=\textwidth]{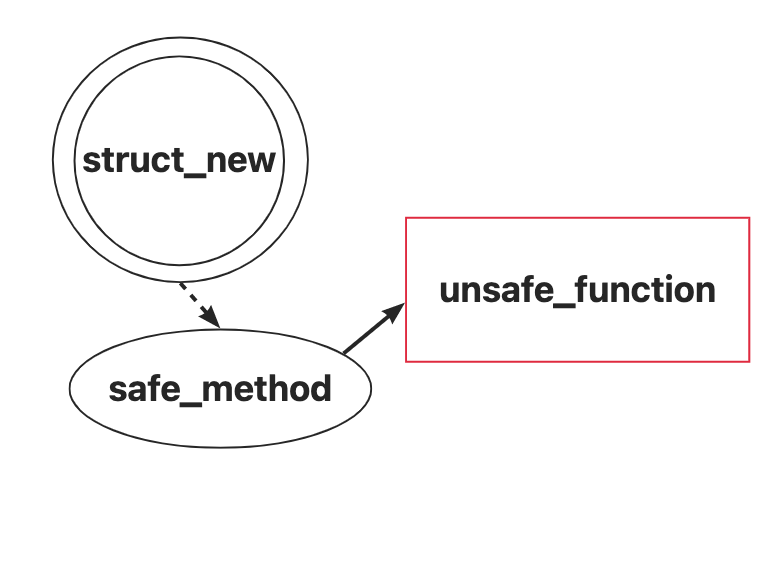}
        \label{fig:sf_um}
        \vspace{-0.6cm}
        \caption{Type I} 
    \end{subfigure}
    \hspace{0.1cm}
    \begin{subfigure}[b]{0.22\textwidth}  
        \includegraphics[width=\textwidth]{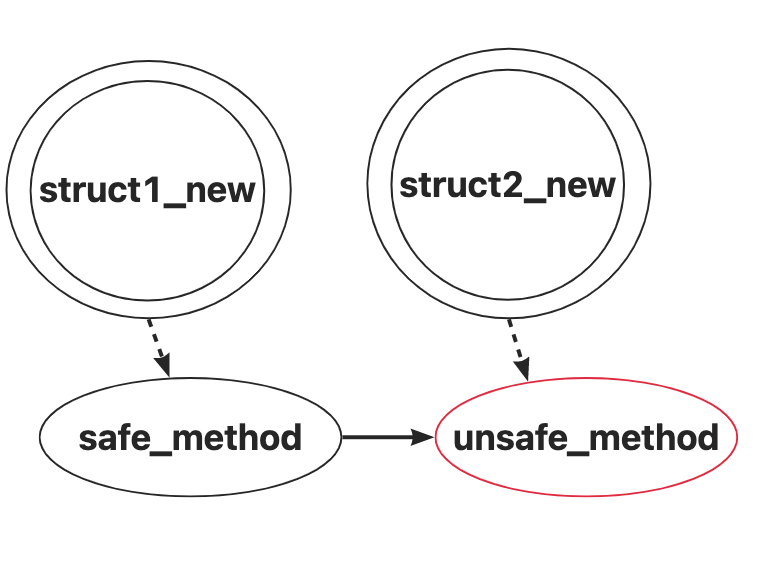}
        \label{fig:sf_uf}
        \vspace{-0.6cm}
        \caption{Type II} 
    \end{subfigure}
    \caption{Indirect Isolation}
\end{figure}

\subsubsection{Indirect Isolation}
This type indicates a scenario where a safe method contains unsafe code, and its constructor is safe. Since the constructor is a necessary precondition for executing the safe method, both the constructor and the method share the responsibility for handling any potential undefined behaviors. There are two subtypes within this category.

\begin{itemize}
    \item $s_m(c_s)\to u_f||c_u$: A safe method $s_m$ calls an unsafe function $u_f$, and the constructor $c_s$ associated with $s_m$ is safe. In such cases, both $c_s$ and $s_m$ bear the responsibility for ensuring the safety requirements of $u_f$ are met. Audit formula: 
    \[RS_{u_f} \subseteq (VS_{s_m} \cup VS_{c_s}) \]
    
    \item $s_m(c1_s)\to u_m(c2_s)$:A safe method $s_m$ calls an unsafe method $u_m$, and the constructors for both $s_m$ and $u_m$ are safe. In this case, the constructor of $s_m$ (denoted as $c1_s$) and $s_m$ itself share the responsibility for ensuring that the safety requirements of $u_m$ are met. Audit formula: 
    \[RS_{u_m} \subseteq (VS_{s_m} \cup VS_{c1_s}) \]
\end{itemize}

\subsubsection{Half Isolation}
This type is similar to the previous two types but also depends on the safety requirements of an unsafe constructor to ensure soundness. Because the unsafe constructor’s practical usage is unpredictable, we refer to this type as half isolation. There are four subtypes within this category.

\begin{figure}[H]
    \begin{subfigure}[b]{0.22\textwidth}
        \includegraphics[width=\textwidth]{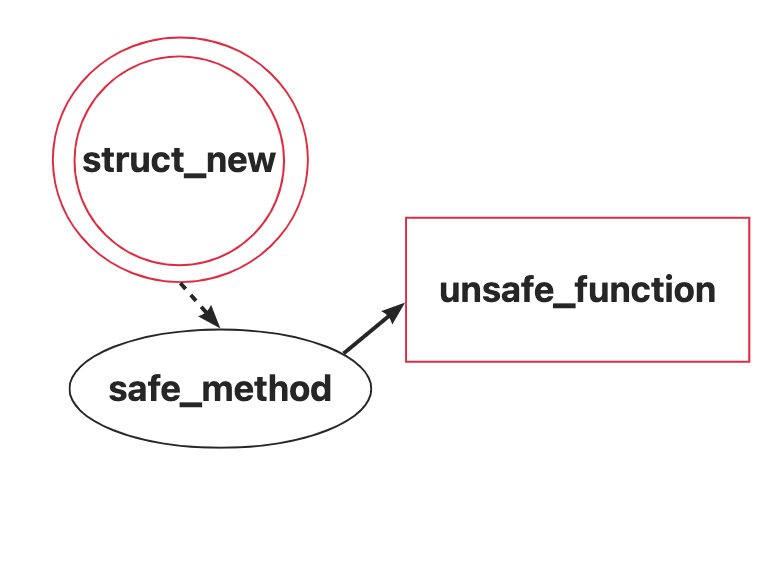}
        \label{fig:sf_um}        
        \vspace{-0.6cm}
        \caption{Type I} 
    \end{subfigure}
    \hfill
    \begin{subfigure}[b]{0.22\textwidth}
        \includegraphics[width=\textwidth]{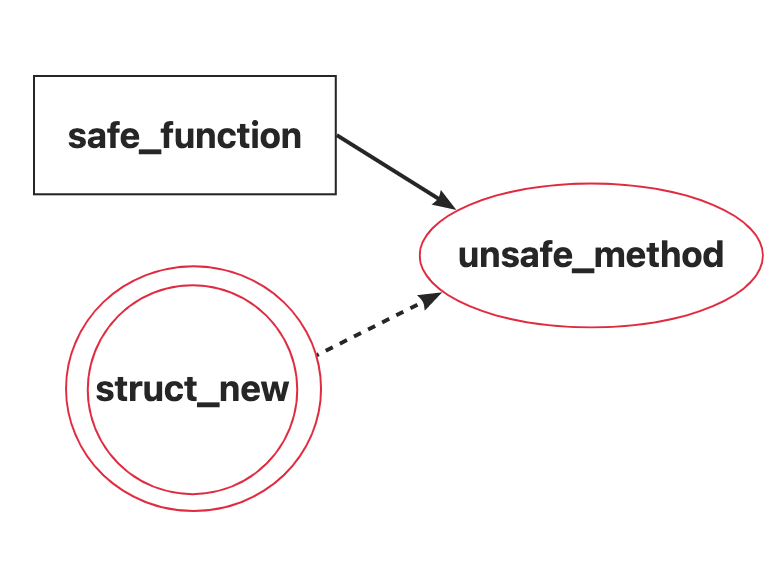}
        \label{fig:sf_um}
        \vspace{-0.6cm}
        \caption{Type II} 
    \end{subfigure}
    \begin{subfigure}[b]{0.22\textwidth}  
        \includegraphics[width=\textwidth]{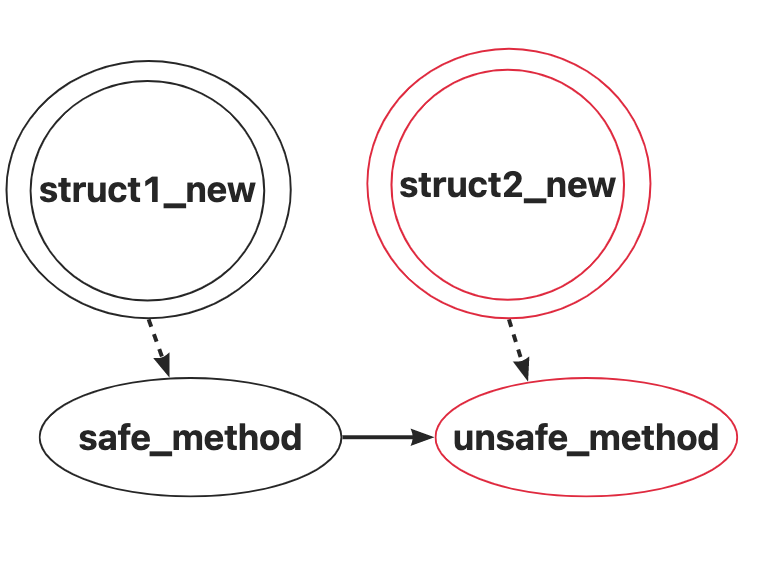}
        \label{fig:sf_uf}
        \vspace{-0.6cm}
        \caption{Type III} 
    \end{subfigure}
    \hfill
    \begin{subfigure}[b]{0.22\textwidth}
        \includegraphics[width=\textwidth]{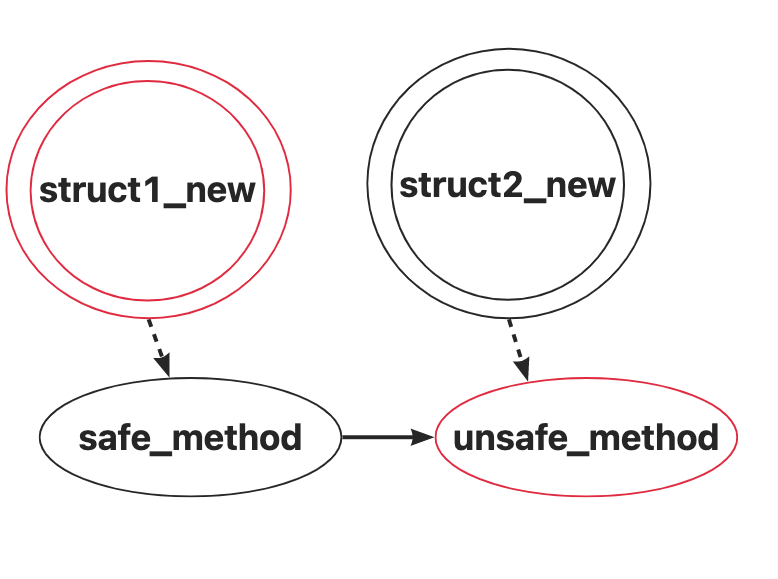}
        \label{fig:sf_um}
        \vspace{-0.6cm}
        \caption{Type IV} 
    \end{subfigure}
     
    \caption{Half Isolation}
\end{figure}

\begin{itemize}
    \item $s_m(c_u)\to u_f||c_u$: In this scenario, a safe method $s_m$ calls an unsafe function $u_f$, and the constructor $c_u$ of $s_m$ is defined as unsafe. In such cases, both $c_u$ and $s_m$ share the responsibility for satisfying the safety requirements of $u_f$. Additionally, since $c_u$ is unsafe and also has safety property requirements $RS(c_u)$, it is assumed that these requirements can be satisfied. Audit formula: 
    \[RS_{u_f} \subseteq (VS_{s_m} \cup RS_{c_u} \cup VS_{c_u})\]
    
    \item $s_f||c_s\to u_m(c_u)$: A safe function $s_f$ calls an unsafe method $u_m$, and the constructor $c_u$ of $u_m$ is unsafe. In such cases, it is considered that $u_m$ should have its own safety requirements independent of $c_u$. This approach allows the verification chains of $u_m$ and $c_u$ to be decoupled from each other. Audit formula: 
    \[RS_{u_m} \subseteq VS_{s_f} \]
    
    \item $s_m(c1_s)\to u_m(c2_u)$: A safe method $s_m$ calls an unsafe method $u_m$, with the constructors of $s_m$ and $u_m$ being safe and unsafe, respectively. Audit formula: 
    \[RS_{u_m} \subseteq (VS_{s_m} \cup VS_{c1_s}) \]
    
    \item $s_m(c1_u)\to u_m(c2_s)$: A safe method $s_m$ calls an unsafe method $u_m$, with the constructors of $s_m$ and $u_m$ being unsafe and safe, respectively. Audit formula: 
    \[RS_{u_m} \subseteq (VS_{s_m} \cup RS_{c1_u} \cup VS_{c1_u}) \]
\end{itemize}

\subsubsection{Open Isolation}
This type of audit unit relies on the safety requirements of two unsafe constructors for verification. It has only one form in our approach.

\begin{figure}[H]
    \includegraphics[width=0.4\linewidth]
    {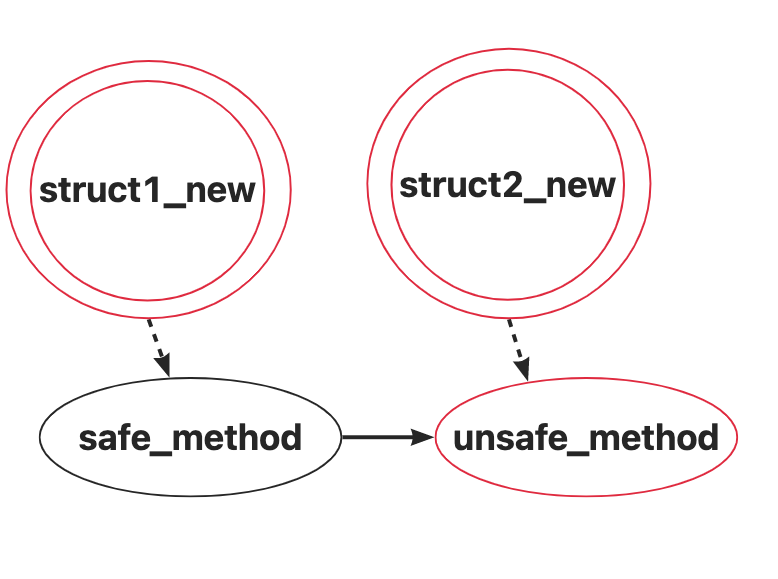}
        \label{fig:sf_um}
        \vspace{-0.5cm}
        \caption{Open Isolation} 
    \label{fig:four_types_of_UIG}
\end{figure}

$s_m(c1_u)\to u_m(c2_u)$: A safe method $s_m$ calls an unsafe method $u_m$, with both $s_m$ and $u_m$ having unsafe constructors. The safety of the execution heavily depends on how the objects are constructed. Audit formula: 
    \[RS_{u_m} \subseteq (VS_{s_m} \cup RS_{c1_u} \cup VS_{c1_u}) \]

The structural patterns provide clear guidance for further safety property verification. This allows our approach to successfully decouple the complicated relationships between safety requirements and verification duties within a Rust crate.

\subsection{Types of Safety Checks}\label{unsafety_bolts}
% characteristics of edges: how to disable unsafety behaviors?
In addition to structural factors, we are also interested in the locations and types of mechanisms that ensure code safety. Intuitively, if unsafe code is encapsulated within a safe function, that function should perform safety verification operations to safeguard the safety of unsafe blocks. We regard these operations as \textit{safety checks}. We categorize safety checks into two types, \textit{direct check} and \textit{no check}, based on the behavior of the unsafe callee's parameters within the safe caller. Our classification does not rely on the specific logic of the project but simply categorizes based on the characteristics of the data flow. 

\textit{Direct Check}: (1) obtained through a global variable, a hardcoded variable, or a global static function returning a fixed value; (2) validated using \textit{if} or \textit{assert} statements

\textit{No Check}: (1) dependent on safe caller's input parameters; (2) lacking any validation of the unsafe callee's input parameters

\section{Experiments}\label{sec:sec_4}
In this section, we design experiments to examine the applicability and usefulness of our approach for real-world Rust project analysis. 
%via UIG, to systematically categorize and scrutinize unsafe code practices within major real-world Rust projects. Through this analysis, we aim to uncover prevalent patterns and identify potential vulnerabilities associated with the use of interior unsafe code. Ultimately, this endeavor seeks to deepen the comprehension and improve the governance of unsafe code in system-level programming contexts.

\subsection{Experimental Setting}
\subsubsection{Rust Project Selection}
For our study, we established two fundamental criteria for selecting projects:
\begin{itemize}
    \item \textit{Maturity and Popularity}: Projects must have been under active development for over two years and maintained within the last year, with a minimum of 1,000 GitHub stars.
    \item \textit{Extensive Use of Unsafe Code}: Projects should contain more than 300 instances of unsafe code.
\end{itemize}

Given these criteria, our research predominantly targets operating system projects due to their extensive engagement with low-level operations and critical focus on safety. This environment is particularly conducive to studying the management and encapsulation of interior unsafe code. We selected four renowned operating system projects within the Rust community:: Asterinas, rCore, Theseus, and Aero.

The Asterinas OS~\cite{asterinas_github}, developed by Ant Group, features the \textit{Framekernel} architecture which emphasizes modularity and security, dividing it into the OS Framework and OS Services.
rCore~\cite{rCore_github} is an educational operating system kernel developed in Rust, demonstrating Rust's capabilities in system programming. 
Theseus~\cite{boos2020theseus} is an innovative operating system that uses language-level mechanisms to enforce safety and correctness, shifting key responsibilities from the OS to the compiler.
The Aero ~\cite{aero_github} is a lightweight Unix-like kernel written in Rust. This OS embraces a monolithic kernel architecture, drawing inspiration from the Linux Kernel.

\subsubsection{Experimental Procedure}
Our experimental methodology employs Algorithm~\ref{upg_generation} to construct each project's unsafe propagation graphs. Following this, we derive unsafe isolation graphs from these UPGs. To validate the applicability and effectiveness of our methods, we implement our audit formulas to each audit unit within the UIGs. This process involves a meticulous manual code review within each unit to detect and categorize instances of unsafe encapsulation and safety checks. The experimental tasks are primarily executed by the first author, with the second author providing a critical review of the statistical outcomes to ensure accuracy. Any discrepancies are resolved through discussion with the third author.

\begin{table*}[t]
\caption{Distribution of unsafe structural types and unsafety checks in four Rust-based projects.}
\label{tab:types_data}
\resizebox{\textwidth}{!}{%
\begin{tabular}{ccclcclcccclcc}
\toprule
\multirow{2}{*}{\textbf{Project}} & \multicolumn{2}{c}{\textbf{Direct Isolation}} &  & \multicolumn{2}{c}{\textbf{Indirect Isolation}} &  & \multicolumn{4}{c}{\textbf{Half Isolation}}                                               &  & \textbf{Open Isolation} & \multirow{2}{*}{\textbf{Total}} \\ \cline{2-3} \cline{5-6} \cline{8-11} \cline{13-13}
                                  & \textbf{sf-uf}      & \textbf{sf-um(sc)}      &  & \textbf{sm(sc)-uf} & \textbf{sm(sc)-um(sc)} &  & \textbf{sm(uc)-uf} & \textbf{sf-um(uc)} & \textbf{sm(sc)-um(uc)} & \textbf{sm(uc)-um(sc)} &  & \textbf{sm(uc)-um(uc)}  &                                 \\ \toprule
Asterinas                         & 96                  & 7                       &  & 43                 & 5                      &  & 11                 & 5                  & 1                      & 19                      &  & 1                       & 188                             \\
rCore                             & 94                  & 14                      &  & 114                & 51                     &  & 0                  & 0                  & 0                      & 0                      &  & 0                       & 273                             \\
Theseus                           & 98                  & 47                      &  & 79                 & 36                     &  & 1                  & 0                  & 0                      & 0                      &  & 0                       & 261                             \\
Aero                           & 158              & 66     &  & 82             & 18                       &  & 4              &  1                  &   0                     &  0                      &  & 4                   & 333                              \\ \bottomrule
\end{tabular}%
}
\end{table*}
\subsection{Characteristics of Unsafety Encapsulation}\label{Characteristics}

\begin{figure}[t]
    \centering
    \includegraphics[width=\columnwidth]{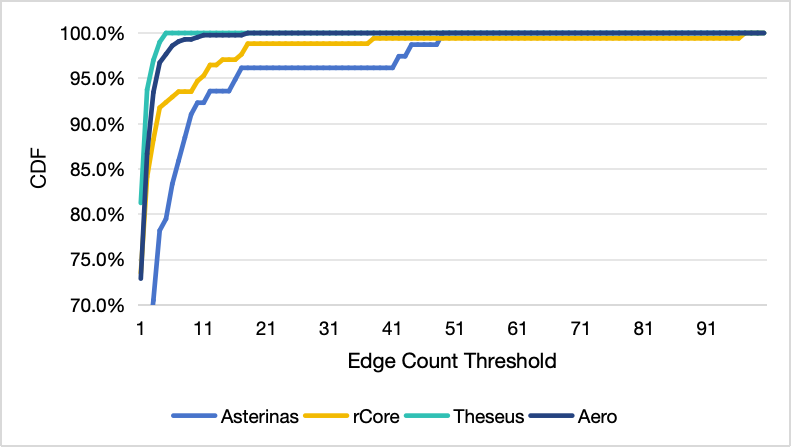}
    \caption{Cumulative distribution of graph sizes based on number of edges in connected subgraphs across four projects.}
    \label{fig:cumulative_graph}
\end{figure}

\subsubsection{Overview of Four Projects}
We implemented our analytical methodology on these four projects, constructing UPGs and deriving UIGs from them. An overview of our statistical findings on audit units within the UIGs is presented in Table ~\ref{tab:types_data}.

A notable trend across all projects is the prevalent use of both \textit{Direct Isolation} and \textit{Indirect Isolation}, which do not incorporate unsafe constructors. On average, unsafe methods constituted 57.8\% of the total audit units, underscoring the necessity for meticulous consideration of several factors, including their constructors, as elaborated in Section~\ref{modeling}. Variations were observed among the projects: rCore, Theseus, and Aero primarily utilize safe constructors for methods involving unsafe code, whereas Asterinas prominently employs unsafe constructors. This is attributed to Asterinas's design, which integrates a high degree of isolation by encapsulating unsafe code within the initialization phases of structures to establish secure interfaces, as confirmed by a developer from Asterinas.

Furthermore, we analyzed each connected subgraph within the UPGs of these projects, with each subgraph delineating the influence scope of the unsafe code it contains. We used the number of edges in each connected graph to reflect the scale of its influence scope. Figure \ref{fig:cumulative_graph} illustrates the cumulative distribution of graph sizes, highlighting differences across projects. In Theseus and Aero, the graphs predominantly reach complete distribution with fewer edges, indicating a direct encapsulation of unsafe code within safe callers and minimal propagation chains. In contrast, Asterinas and rCore respectively have 20.5\% and 7.6\% of their connected graphs containing more than five edges, involving 53.0\% and 35.0\% of all nodes. This suggests extensive unsafety propagation involving multiple unsafe nodes and dependencies on constructors in these projects, highlighting a complex landscape of unsafe code management.

In summary, The deployment of interior unsafe code in real-world projects is characterized by a high level of complexity, particularly due to the significant involvement of constructors. This often leads to intricate and interdependent call relationships.
% Not all unsafe APIs are directly encapsulated within safe APIs. In the UPGs we generated, each function call is represented as an edge. The proportions of unsafe-to-unsafe calls to total call edges are 28.5\% for Asterinas, 32.0\% for rCore, and 39.8\% for Theseus. Figure ~\ref{fig:uu_su_comparison} illustrates the comparison between unsafe-to-unsafe and safe-to-unsafe call frequencies across these three projects. \textit{Direct Encapsulation}(safe function calls unsafe function) and \textit{Single Encapsulation}(safe method calls unsafe function) are the predominant types of unsafe encapsulation across the three projects.

% Table ~\ref{tab:types_data} provides statistics on the origin of the definitions for the called unsafe functions or methods. Regarding the unsafe callees that are custom-defined within the crate compared to those originating from external libraries such as the standard library or third-party crates, aside from Theseus, which exhibits a disparity of 26.5\%, there is no significant difference between Asterinas(1.0\%) and rCore(5.2\%).

\subsubsection{Complexity of Different Audit Unit}
This subsection illustrates the complexity involved in verifying different types of audit units through specific cases encountered within these projects. We will also demonstrate the powerful decoupling capability of our method for validating the encapsulation of interior unsafe code.

For \textit{Direct Isolation}, the encapsulation of unsafe code is relatively explicit. Developers and code auditors must identify all preconditions and postconditions associated with the unsafe code. The main verification task is to ensure these safety prerequisites are comprehensively addressed within the internal logic of the safe caller, thereby affirming the soundness of the encapsulation.

In \textit{Indirect Isolation}, the safe caller is available following construction by a safe constructor. Our audit formula stipulates that the combined safety guarantees from both the safe caller and its constructor must encompass the immediate safety needs of the unsafe callee. For instance, as detailed in Listing \ref{lst:XApic}, the unsafe callee \texttt{read\_volatile} from the standard library mandates that its parameters should be \textit{valid} and \textit{aligned} at the syntactic level. These requirements are proactively managed in \texttt{XApic}'s constructor \texttt{new}, which performs essential initialization to ensure the \texttt{mmio\_region} field is correctly prepared. Subsequently, the safe method caller \texttt{read} can only conduct project-specific validations and then directly use this field as a parameter to invoke the unsafe callee \texttt{read\_volatile}.

The verification challenges of \textit{Half Isolation} and \textit{Open Isolation} lies in their reliance on comprehensive encapsulation across multiple audit units. Unlike \textit{Direct Isolation} and \textit{Indirect Isolation}, which encapsulate the safety of unsafe code through the behavior of a single audit unit, \textit{Half Isolation} and \textit{Open Isolation} necessitate multi-hop verifications. In listing ~\ref{lst:IoPort}, 
the safety of safe method \texttt{read}, ensured by its constructor, is contingent on the known integrity of the I/O port. If the I/O port's identity is unclear, it may pose significant memory safety risks. Within this audit unit, we cannot judge how its unsafe constructor is used, making it challenging to ensure the validity of its field \texttt{port}. This introduces a complexity in verification that requires examining another audit unit, where another caller invokes \texttt{IoPort::new}, to verify its safety before confirming the safety of the \texttt{read} method. This necessitates a two-hop verification for \textit{Half Isolation}. Similarly, Open Isolation may demand two or three hops for thorough verification, depending on the interconnectedness of the components and their respective constructors.

\begin{lstlisting}[language=rust, caption={Half Isolation: sm(uc)-uf from Asterinas}, label={lst:IoPort}]
impl<T, A> IoPort<T, A> {
    /// # Safety: creating an I/O port is a privileged operation.
    pub const unsafe fn new(port: u16) -> Self {
        Self { port, ... }
    }
    pub fn read(&self) -> T {
        unsafe { PortRead::read_from_port(self.port) }
    }
}
\end{lstlisting}

\noindent\begin{minipage}{\linewidth}
\begin{lstlisting}[language=rust, caption={Indirect Isolation: sm(sc)-uf from Asterinas}, label={lst:XApic}][ht]
pub struct XApic {
    mmio_region: &'static mut [u32],
}
impl XApic {
    pub fn new() -> Option<Self> {
        ...
        let address = vm::paddr_to_vaddr(get_apic_base_address());
        let region: &'static mut [u32] = 
            unsafe { &mut *(address as *mut [u32; 256]) };
        Some(Self { mmio_region: region })
    }
    fn read(&self, offset: u32) -> u32 {
        assert!(offset as usize % 4 == 0);
        let index = offset as usize / 4;
        unsafe { ptr::read_volatile(&self.mmio_region[index]) }
    }
}
\end{lstlisting}
\end{minipage}

\subsubsection{Discussion of Struct with Both Safe and Unsafe Constructors}
In our previous discussions, we examined scenarios where structs were characterized by either a single safe or unsafe constructor. A significant advantage of the UIG model we propose is its flexibility in managing complex structures, particularly those equipped with multiple constructors. This capability is crucial as it allows a large struct to be systematically decomposed into several audit units, thereby facilitating the meticulous handling of scenarios involving multiple constructors.

\begin{figure}[h]
    \centering
    \includegraphics[width=\columnwidth]{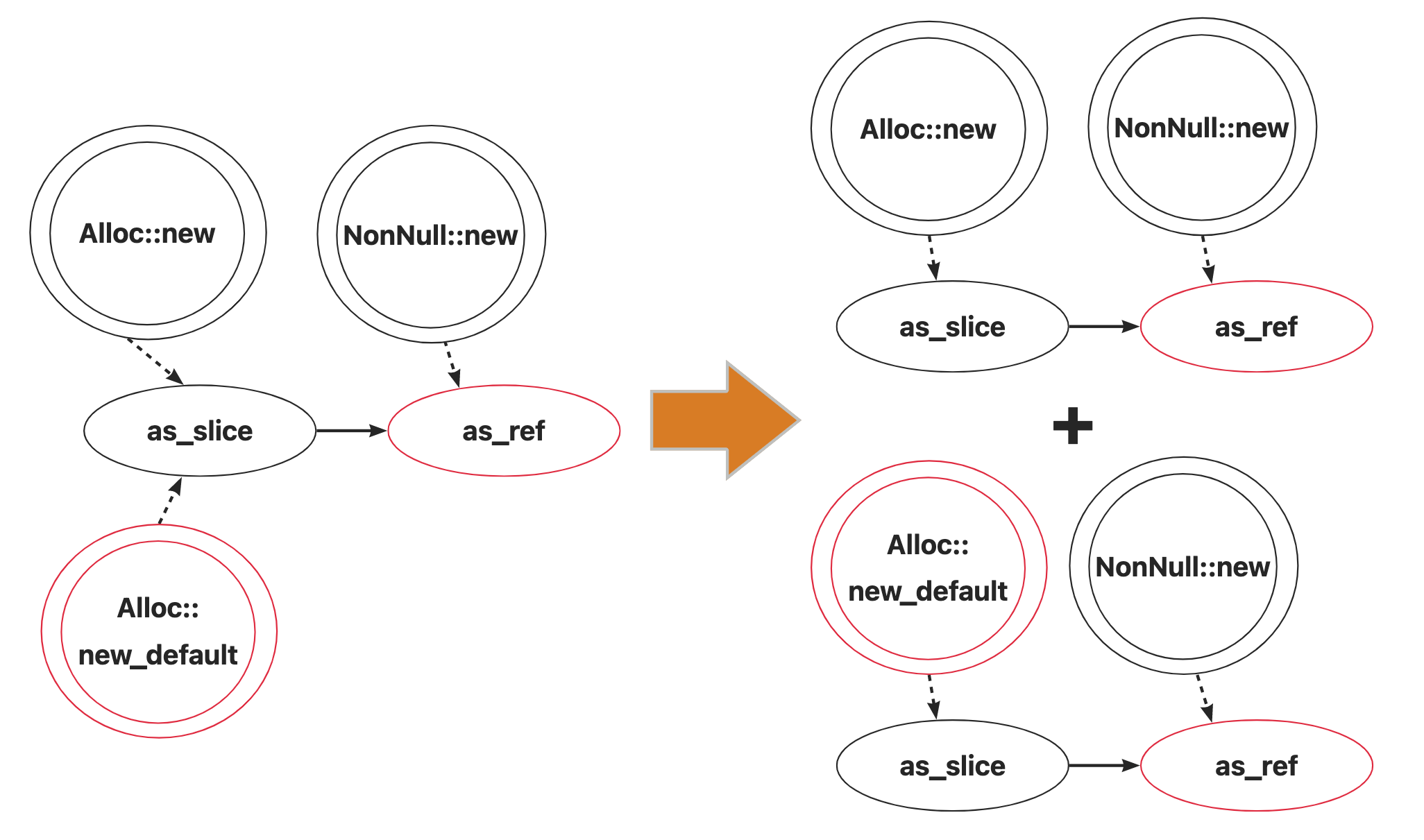}
    \caption{Decomposition of a large structure with both safe and unsafe constructors into multiple audit units.}
    \label{fig:sc_uc_combination}
\end{figure}

When a struct incorporates both safe and unsafe constructors, the analysis involves segregating these constructors into distinct audit units. This allows for two primary pathways to utilize this method. No matter which way is taken, the usage of the methods should be sound. If a struct is instantiated via its safe constructor, the methods that use interior unsafe code should be directly usable. Otherwise, they must be marked as unsafe. This is also the reason we discussed in Chapter ~\ref{sec:method} for why the modeling of audit unit is considered reasonable. As for the construction utilizing an unsafe constructor, if its safety annotations are strictly satisfied, then the program state should be consistent with that obtained through a safe constructor. Consequently, the scope of safety covered by an unsafe constructor, when combined with its specific safety requirements, must align with the safety coverage provided by the safe constructor. This alignment ensures that regardless of the construction method used, the structural integrity and safety standards remain consistent and reliable. That means: 
\[VS_{c_s} \equiv VS_{c_s} \cup RS_{c_s}\]

Therefore, for every safe method $m_i$ in it, if $m_i$ calls an unsafe function $u_f$ or unsafe method $u_m$, audit formula can be expressed as follows:
\[RS_{u_f\ or\ u_m} \subseteq (VS_{c_s} \cap (VS_{c_u} \cup RS_{c_u})) \cup VS_{m_i} \]

In our experiments, we have identified several potentially problematic examples through our methodology.
Listing \ref{lst:sc and uc in Aero} displays a code snippet from \texttt{Aero}. 
The \texttt{PhysAddr} struct incorporates a safe constructor \texttt{new}, which performs necessary checks on the parameter \texttt{addr} through \texttt{assert\_eq!}, and an unsafe constructor \texttt{new\_unchecked}, which directly invokes \texttt{PhysAddr} to generate a physical address. Therefore, based on their program behaviors, we can construct their verified safety sets: $VS_{new}$ = \{ Init\_State \}, and $VS_{new\_unchecked}$ = $\emptyset$. To maintain object consistency, the unsafe constructor should be annotated to indicate that parameter checks are still necessary. So we get its required safety set $RS_{new\_unchecked}$ = \{ Init\_State \}. This demonstrates the equivalence between the two construction methods. These sets facilitate the application of our audit formula to evaluate the safety of methods within \texttt{PhysAddr}. For example, the safe method \texttt{as\_hhdm\_virt} employs a global mutable static variable, which requires an unsafe marker as this could lead to data races and other thread safety issues. These issues primarily arise from the global program state, which falls outside the safety guarantees provided by both constructors according to our audit formula. Consequently, labeling it as 'safe' may lead to unsound problems. This concern is corroborated by developer comments.
% be initialized so that we can establish the set: $RS_{NonNull::as\_ref}$ = \{ Allocated \}. It meets the safety criteria when derived from both constructors, affirming its safe encapsulation. Besides, the unsafe method \texttt{as\_ptr\_mut} using a mutable raw pointer, requires the addressed location to be initialized. Thus, we can generate: $RS_{as\_ptr\_mut}$ = \{ Initialized, Allocated \}. It can also pass our audit formula. This method's $RS$ is somewhat coupled with its unsafe constructor, and being marked as unsafe may reflect the developers' more cautious considerations.

\noindent\begin{minipage}{\linewidth}
\begin{lstlisting}[language=rust, caption={case sf-um that has both safe and unsafe constructor from Aero}, label={lst:sc and uc in Aero}]
impl PhysAddr {
    /// # Safety: Bits in the range 52 to 64 have requirements.
    //unsafe constructor
    pub const unsafe fn new_unchecked(addr: u64) -> PhysAddr {
        PhysAddr(addr)
    }
    //safe constructor
    pub fn new(addr: u64) -> PhysAddr {
        assert_eq!( addr.get_bits(52..64), 0,
            "Can not have any bits in the range 52 to 64 set"
        );
        unsafe { PhysAddr::new_unchecked(addr) }
    }
    //safe method
    pub fn as_hhdm_virt(&self) -> VirtAddr {
        // TODO: Make `PHYSICAL_MEMORY_OFFSET` an atomic usize 
        // instead of making it a mutable static and spamming 
        // `unsafe` everywhere; where its not even required.
        unsafe { crate::PHYSICAL_MEMORY_OFFSET + self.as_u64() }
    }
}
\end{lstlisting}
\end{minipage}

\begin{table}[h]
\centering
\caption{Status of safety annotations for custom unsafe functions within the projects. The table presents the comparison of functions lacking annotations to the total number of unsafe functions.}
\label{tab:lack_annotation}
\begin{tabular}{cccc}
\toprule
\multirow{2}{*}{\textbf{Project}} & \multicolumn{3}{c}{\textbf{Annotation(lack/total)}} \\ \cline{2-4} 
                         & \textbf{Constructor}       & \textbf{Function}       & \textbf{Method}       \\ \toprule
Asterinas                & 1/20        &  9/27    & 12/24  \\
rCore                    & 10/10       & 58/58    & 30/30  \\
Theseus                  & 8/8         & 83/95    & 23/37  \\ 
Aero                  & 4/7         & 53/57    & 36/47  \\
\bottomrule
\end{tabular}
\end{table}

\section{Issues and Findings}
\subsection{Lack of Safety Annotations}
\uline{\textit{Issue 1}: Many unsafe functions or methods lack safety requirement annotations, which, while not strictly necessary, are very useful for achieving safety encapsulation.}
% The lack of safety annotations on unsafe constructors can have a wider impact and significantly increase the complexity of verification.

% In Rust project development, unsafe code typically needs to be annotated with its safety requirements to alert other developers about the considerations they need to keep in mind. This issue is particularly crucial when dealing with \textit{Half Isolation} and \textit{Open Isolation}.
Within the Rust community, it is a common practice to annotate custom unsafe functions with required safety requirements immediately after developing them~\cite{unsaferust}. This practice ensures that the safety implications of using such functions are clear to all developers involved. However, our empirical findings reveal that this paradigm is not universally adhered to, primarily because it is not a mandatory requirement. In Table ~\ref{tab:lack_annotation}, we have listed the safety annotation status of all custom unsafe functions in these four projects. Asterinas, developed by a company, exhibits a higher density of safety annotations compared to other projects with a 31\% absence rate. However, this rate in the other three projects reaches as high as 87\%, as they are presumably with less stringent development protocols.

For \textit{Direct Isolation} and \textit{Indirect Isolation}, the lack of safety annotations within their functions and methods necessitates that developers and code reviewers need to manually audit and extract the safety requirements of the unsafe callee, and then assess whether the safe caller's encapsulation behavior satisfies these safety requirements. 

However, the absence of safety annotations on unsafe constructors can significantly undermine the auditing principle that typically focuses only on the nodes within one single audit unit. This shortfall is critical because unsafe constructors often fail to provide explicit safety requirements in their code. Consequently, these constructors necessitate a broader examination that extends beyond their immediate context to include the behavior of their callers in other audit units. Listing ~\ref{lst:local_key} provides an example of \textit{Half Isolation} from Theseus. To confirm that this unsafe encapsulation is sound, the sets \textit{$RS_{inner}$} and \textit{$RS_{new}$} are necessary. While the construction logic in line 6 obscures the clarity of its safety requirements to build the \textit{$RS_{new}$}. This complexity underscores the need for a broader investigation of its safety. Additionally, as unsafe constructors serve as boundaries in the audit unit of \textit{Half Isolation} and \textit{Open Isolation}, the absence of their safety can impact multiple other audit units.

\begin{figure}[t]
    \centering
    \includegraphics[width=\columnwidth]{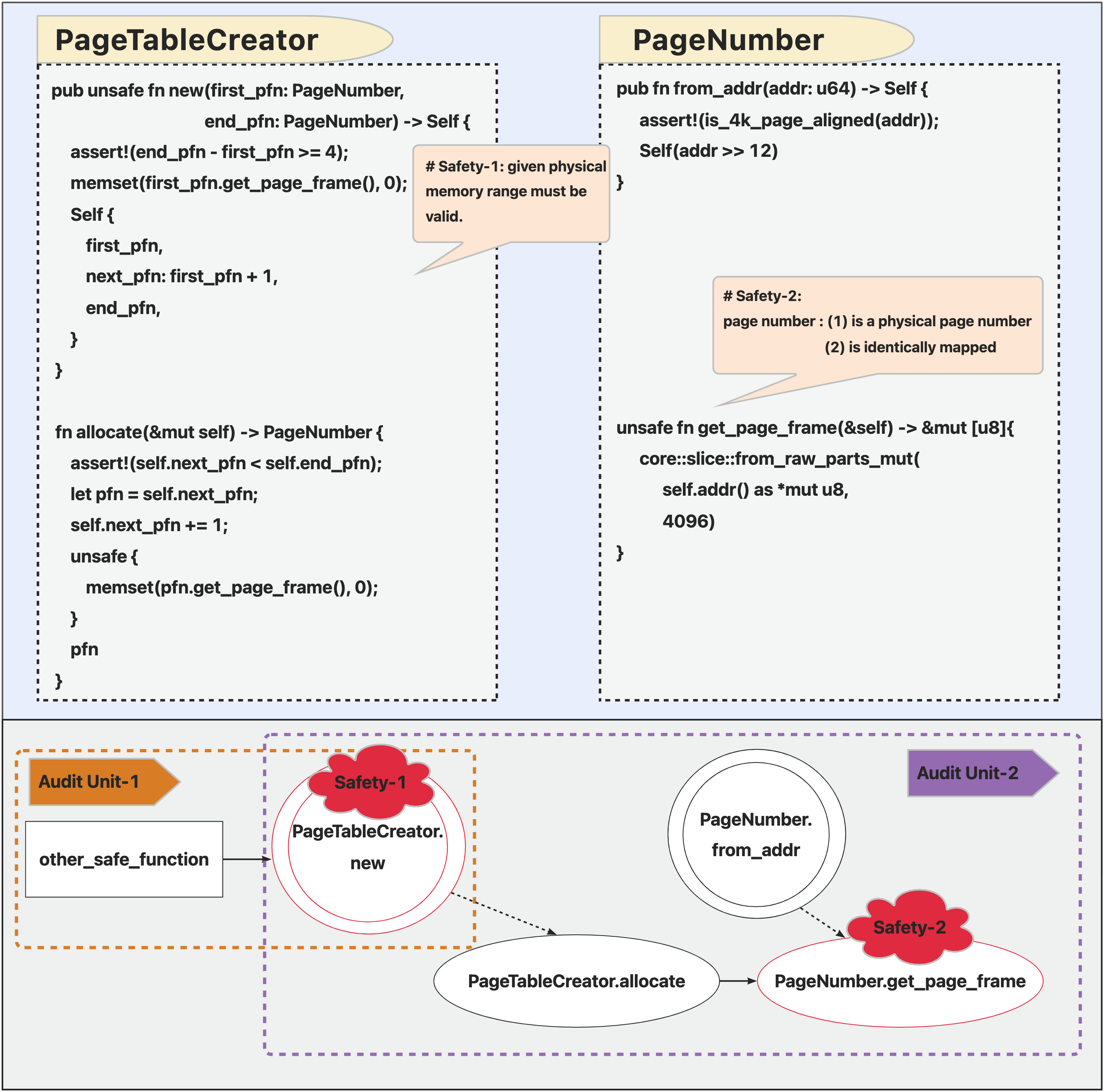}
    \caption{An example of a half isolation case: Annotating the unsafe constructor with safety requirements can simplify verification by treating it as two independent audit units.}
    \label{fig:pagetablecreator}
\end{figure}

Another case from Asterinas provides a best practice in Figure ~\ref{fig:pagetablecreator}. In the example, the $RS$ set of the unsafe callee \texttt{get\_page\_frame} can be easily obtained from annotations: page number (1) is a physical page number, and (2) is identically mapped. Meanwhile, code reviewers only need to focus on the code within the \texttt{PageTableCreator} to obtain sets \textit{$RS_{new}$} and \textit{$VS_{new}$}. In our method, within this audit unit, \textit{$RS_{new}$} is assumed to always be met. Its assurance of safety is evaluated within a distinct audit unit. Therefore, even if the \textit{$VS$} of the safe caller \texttt{allocate} is not sufficient to cover all safety of get\_page\_frame, the \textit{$RS_{new}$} can assist in confirming the validity of the PageNumber object. 

\noindent\begin{minipage}{\linewidth} 
\begin{lstlisting}[language=rust, caption={Half isolation: sm(uc)-uf from Theseus}, label={lst:local_key}]
pub struct LocalKey<T: 'static> {
    inner: unsafe fn() -> T,
}
impl<T: 'static> LocalKey<T> {
    pub const unsafe fn new(inner: unsafe fn()->T) -> LocalKey<T> {
        LocalKey { inner }
    }
    pub fn try_with<F, R>(&'static self, f: F){
        unsafe { let thread_local = (self.inner)();}
    }
}
\end{lstlisting}
\end{minipage}

In summary, audit units help identify critical areas in the development of unsafe code, streamlining the verification process with structured audit formulas. However, our findings highlight a significant shortfall in safety annotations across major Rust-based operating system projects, complicating verification and increasing the risk of safety oversights.

Based on the \textit{issue 1} we have identified, we propose the following suggestions for project developers:

\uline{\textit{Suggestion 1}: Developers should annotate each unsafe function and method with explicit safety requirements.
The Rust compiler could enforce such checks, regardless of their semantic correctness.}%Ensure each safety annotation is clear enough to construct safety set and assess unions, intersections, and subsets.}

\subsection{Unsoundness Concerning Visibility}
\uline{\textit{Issue 2}: Developers may overlook the soundness issues in certain internal module scenarios, such as bypassing unsafe constructors via literal constructors or directly invoking some safe methods that lack safety checks.}
\subsubsection{Literal Construction}
\uline{Rust's literal construction can bypass necessary safety assurances in constructors, which may potentially compromise the methods' soundness.}
In Rust, structs can be instantiated in two primary ways: using literal syntax and constructors. 
Literal syntax directly sets a structure's fields if they are accessible. Constructors, on the other hand, aim for uniform initialization of the structure's fields.
Previously, we discussed that objects created via safe or annotated unsafe constructors must maintain a consistent program state for predictable behaviors. However, using literal construction can jeopardize this consistency, as it allows direct field manipulation.

\begin{figure}[t]
    \centering
    \includegraphics[width=0.9\columnwidth]{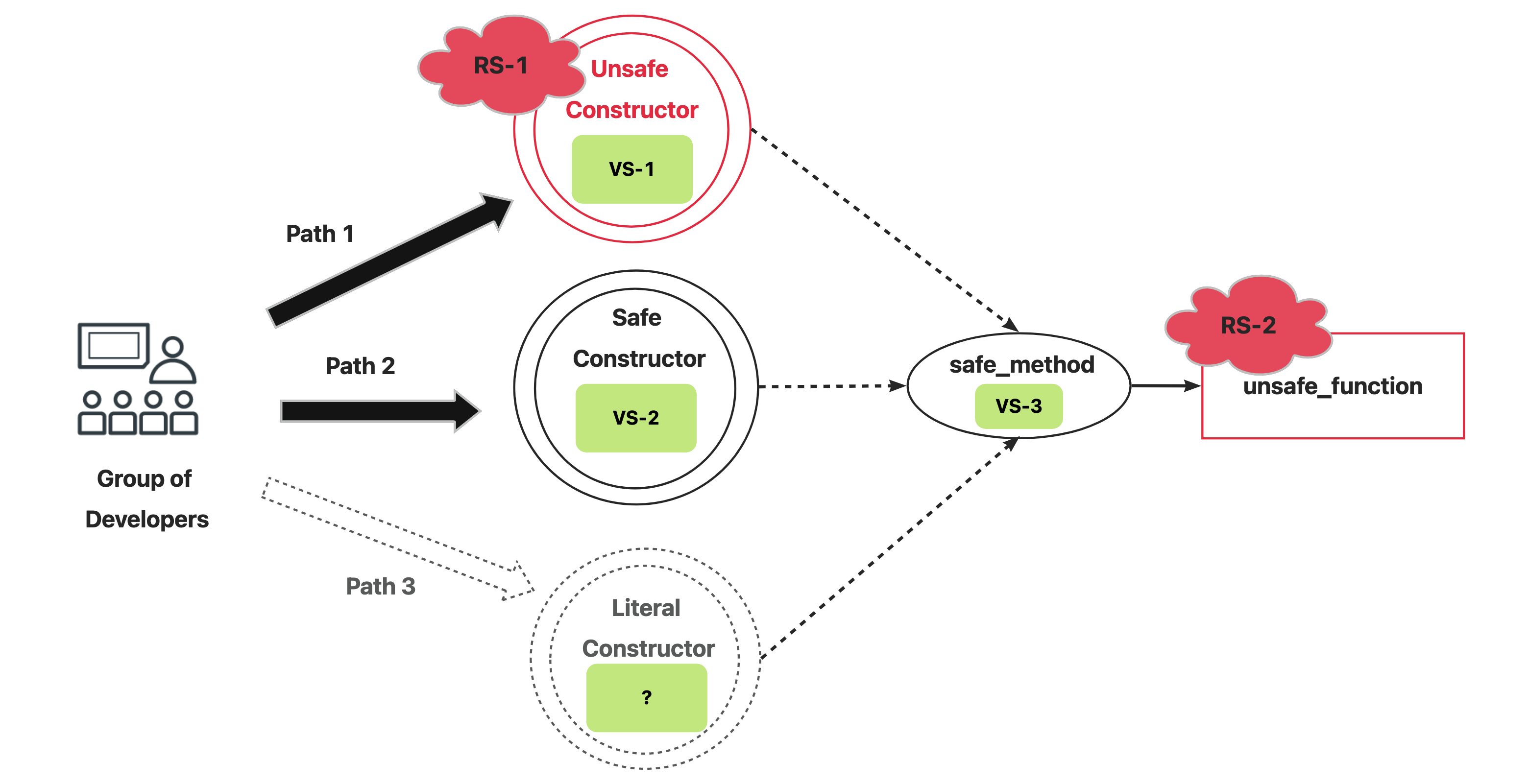}
    \caption{A model depicting the potential unsoundness introduced by literal construction in Rust, which may bypass safety guarantees provided by constructors.}
    \label{fig:literal}
\end{figure}

We model the behavior associated with literal constructors as an issue in Figure ~\ref{fig:literal}. 
In this scenario, developers in a team have three paths to access a struct's method. If the unsafe constructor's safety annotations are sufficient, path one is a viable choice since developers can recognize potential safety issues. Choosing path 2 means the safe constructor inherently encapsulates the unsafety. However, literal construction obscures these details. If other developers in the team attempt to use the struct's \texttt{safe\_method} interface via literal construction, they might not establish a comprehensive \textit{VS} for this method. This could result in the \texttt{safe\_method} triggering an unsoundness issue.
Although this issue might be partly attributed to non-standard development practices, we want to emphasize that this is a potential flaw in Rust's current development of interior unsafe code.
In real-world projects, numerous instances exist where constructors encapsulate methods' unsafety, yet literal construction could potentially leave a vulnerability in such code, thus becoming a possible form of attack against Rust's safety guarantees.

Regarding this issue, we would like to make suggestions to the developers of the Rust compiler:

\uline{\textit{Suggestion 2}: The Rust compiler may consider disabling literal constructors in the future to better support unsafe constructors and ensure soundness.}

\subsubsection{Functions with No Safety Check}
\uline{In the development practices, developers may opt to designate code that lacks safety checks as non-public.}

In our experiment results, a counterintuitive observation was the presence of \textit{no blot} data, where safe callers employ unsafe code without any safety checks on the input parameters. As detailed in Table~\ref{tab:bolts_data}, the data of no safety checks, on average, 69.0\% are related to methods. In Section~\ref{Characteristics}, our discussions on the structural properties of audit units can explain why it is reasonable for methods to exhibit \textit{no blot} interior unsafe code, as their constructors can provide a layer of safety assurance. However, the scenario for static functions proved perplexing. Some examples may ensure safety through global states or parameters, while others still have safety issues to be addressed, such as the code case in listing ~\ref{lst:sc and uc in Aero}. Our further analysis showed that 41.9\% of the function's \textit{no blot} occurrences are related to visibility. Listing~\ref{visibility} from Theseus illustrates that a static function directly dereferences a raw pointer after accepting it as an input parameter. When the parameter is a null pointer or other invalid values, this action may trigger an exception. Since this function's usage is confined within its module, it is controllable to ensure the soundness of its parameters among its limited callers. This may be a common development practice, yet it somewhat contradicts the principle of encapsulating interior unsafe code.

\uline{\textit{Suggestion 3}: Developers should double-check that their interior unsafe code, which lacks safety checks, is sound. The Rust compiler might issue a warning in such cases.}

\begin{table}[t]
\caption{Statistics of cases with no safety checks.}
\label{tab:bolts_data}
\resizebox{0.9\columnwidth}{!}{%
\begin{tabular}{cccccccccccc}
\toprule
\multirow{2}{*}{\textbf{Isolation Type}} & \multicolumn{2}{c}{\textbf{Asterinas}} &  & \multicolumn{2}{c}{\textbf{rCore}} &  & \multicolumn{2}{c}{\textbf{Theseus}} &  & \multicolumn{2}{c}{\textbf{Aero}} \\ \cline{2-3} \cline{5-6} \cline{8-9} \cline{11-12} 
                                         & No                & Yes                &  & No               & Yes             &  & No               & Yes               &  & No               & Yes               \\ \toprule
sf-uf                           & 28                & 74                 &  & 15               &  94             &  & 31               &   69                &  &     28             &   161                \\
sf-um(sc)                       & 2                 & 4                  &  & 2                &   14            &  & 7               &   40                &  &       18           &   71                \\
sm(sc)-uf                       & 23                & 20                 &  & 49              &  88               &  & 37              &  43                 &  &     18             &  79                 \\
sm(sc)-um(sc)                   & 0                 & 5                  &  & 14               & 47                &  & 16               & 26                  &  &   15               &  71                 \\
sm(uc)-uf                       & 11                & 0                  &  & 0                & 0                &  & 0                &  1                 &  &     0             &      5             \\
sf-um(uc)                       & 1                 & 4                  &  & 0                &  0               &  & 0                &  0                 &  &     0             &     2              \\
sm(sc)-um(uc)                   & 1                 & 0                  &  & 0                &   0              &  & 0                &   0                &  &     4             &     3              \\
sm(uc)-um(sc)                   & 9                 & 13                 &  & 0                &    0             &  & 0                &    0               &  &     0             &    0               \\
sm(uc)-um(uc)                   & 0                 & 1                  &  & 0                &     0            &  & 0                &     0              &  &     0             &    5               \\ \hline
\textbf{Total} & \textbf{75} & \textbf{121}  &  & \textbf{80}  & \textbf{241}  &  & \textbf{91} & \textbf{179} &  & \textbf{83}  & \textbf{397}  \\ \bottomrule
\end{tabular}%
}
\end{table}

\noindent\begin{minipage}{\linewidth} 
\begin{lstlisting}[language=rust, caption={A sample case with no safety checks from Theseus}, label={visibility}]
fn panic_callback(data_ptr: *mut u8, ...) {
    let data = unsafe { &mut *(data_ptr as *mut TryIntrinsicArg) };
    ...
}
\end{lstlisting}
\end{minipage}

\section{Related Work}
In this section, we discuss and compare with investigations on unsafe Rust code previous to our work to show the significance of our approach.

Researches in unsafe code isolation focus on strategies that leverage both programming language features~\cite{burtsev2021isolation,narayanan2020redleaf} and system-level controls~\cite{rivera2021keeping, almohri2018fidelius, liu2020securing} to secure memory and data flow. These methods consider the isolation of unsafe code from a runtime perspective, rather than during development and auditing.

Formal verification~\cite{jung2017rustbelt,matsushita2022rusthornbelt} is a robust yet complex method for validating the safety of Rust's unsafe code, but it can only verify programs at the language level. In contrast, our method also effectively aids developers in auditing the logical safety of custom unsafe functions. 

Static analysis methods leverage Rust's strong type system and ownership model to provide precise and accessible bug detection. They aim to identify common bug patterns in unsafe Rust code, such as memory deallocation violations~\cite{cui2023safedrop, li2021mirchecker}, panic safety bugs, higher-order invariant bugs, and Send/Sync variance bugs~\cite{bae2021rudra}.

Empirical studies give insights into the unsafe Rust usage status~\cite{mccormack2024against,yu2019fearless,fulton2021benefits} and point out why and how to write unsafe code in real-world programs~\cite{evans2020rust,astrauskas2020programmers}. Researchers also classify the unsafe APIs in standard libraries based on safety descriptions~\cite{cui2024unsafe}. However, they focus more on the unsafe code itself and don't take a deeper look at how to securely encapsulate unsafe code in safe Rust. 

\section{Conclusion}
% target
In this paper, our research aimed to identify common patterns and potential vulnerabilities in the encapsulation of unsafe code in Rust.
% method
We developed graph-based models called the unsafety propagation graph and the unsafety isolation graph to systematically analyze and characterize unsafe coding practices. We further identified nine categories of audit units.
% application
Then we applied our methodologies to four Rust-based operating system projects. The experiment results confirmed the complexity of unsafe code encapsulation and demonstrated the applicability and effectiveness of our method in auditing interior unsafe code.
% finding
Moreover, we also identified two types of issues prevalent in current Rust projects: the lack of annotations on custom unsafe functions and the soundness problems introduced by literal constructions and non-public functions. 
% looking ahead
Through these efforts, we believe our method will promote the standardization of unsafe code development and auditing within the Rust community.

\bibliographystyle{ACM-Reference-Format}
\bibliography{ref.bib}


\begin{thebibliography}{28}

%%% the Web address.

\ifx \showCODEN    \undefined \def \showCODEN     #1{\unskip}     \fi
\ifx \showDOI      \undefined \def \showDOI       #1{#1}\fi
\ifx \showISBNx    \undefined \def \showISBNx     #1{\unskip}     \fi
\ifx \showISBNxiii \undefined \def \showISBNxiii  #1{\unskip}     \fi
\ifx \showISSN     \undefined \def \showISSN      #1{\unskip}     \fi
\ifx \showLCCN     \undefined \def \showLCCN      #1{\unskip}     \fi
\ifx \shownote     \undefined \def \shownote      #1{#1}          \fi
\ifx \showarticletitle \undefined \def \showarticletitle #1{#1}   \fi
\ifx \showURL      \undefined \def \showURL       {\relax}        \fi
% invisible to TeX
\providecommand\bibfield[2]{#2}
\providecommand\bibinfo[2]{#2}
\providecommand\natexlab[1]{#1}
\providecommand\showeprint[2][]{arXiv:#2}

\bibitem[rus(2024)]%
        {rustfunctions}
 \bibinfo{year}{2024}\natexlab{}.
\newblock \bibinfo{title}{Functions - The Rust Programming Language}.
\newblock
\newblock
\urldef\tempurl%
\url{https://doc.rust-lang.org/reference/items/functions.html}
\showURL{%
\tempurl}
\newblock
\shownote{[Accessed: 2024-06-03]}.


\bibitem[{Aero Contributors}(2024)]%
        {aero_github}
\bibfield{author}{\bibinfo{person}{{Aero Contributors}}.} \bibinfo{year}{2024}\natexlab{}.
\newblock \bibinfo{title}{{Aero}}.
\newblock
\newblock
\urldef\tempurl%
\url{https://github.com/Andy-Python-Programmer/aero}
\showURL{%
\tempurl}
\newblock
\shownote{[Accessed: 2024-05-20]}.


\bibitem[Almohri and Evans(2018)]%
        {almohri2018fidelius}
\bibfield{author}{\bibinfo{person}{Hussain~MJ Almohri} {and} \bibinfo{person}{David Evans}.} \bibinfo{year}{2018}\natexlab{}.
\newblock \showarticletitle{Fidelius charm: Isolating unsafe rust code}. In \bibinfo{booktitle}{\emph{Proceedings of the Eighth ACM Conference on Data and Application Security and Privacy}}. \bibinfo{pages}{248--255}.
\newblock


\bibitem[Astrauskas et~al\mbox{.}(2022)]%
        {astrauskas2022prusti}
\bibfield{author}{\bibinfo{person}{Vytautas Astrauskas}, \bibinfo{person}{Aurel B{\'\i}l{\`y}}, \bibinfo{person}{Jon{\'a}{\v{s}} Fiala}, \bibinfo{person}{Zachary Grannan}, \bibinfo{person}{Christoph Matheja}, \bibinfo{person}{Peter M{\"u}ller}, \bibinfo{person}{Federico Poli}, {and} \bibinfo{person}{Alexander~J Summers}.} \bibinfo{year}{2022}\natexlab{}.
\newblock \showarticletitle{The prusti project: Formal verification for rust}. In \bibinfo{booktitle}{\emph{NASA Formal Methods Symposium}}. Springer, \bibinfo{pages}{88--108}.
\newblock


\bibitem[Astrauskas et~al\mbox{.}(2020)]%
        {astrauskas2020programmers}
\bibfield{author}{\bibinfo{person}{Vytautas Astrauskas}, \bibinfo{person}{Christoph Matheja}, \bibinfo{person}{Federico Poli}, \bibinfo{person}{Peter M{\"u}ller}, {and} \bibinfo{person}{Alexander~J Summers}.} \bibinfo{year}{2020}\natexlab{}.
\newblock \showarticletitle{How do programmers use unsafe \text{Rust}?}
\newblock \bibinfo{journal}{\emph{Proceedings of the ACM on Programming Languages}} \bibinfo{volume}{4}, \bibinfo{number}{OOPSLA} (\bibinfo{year}{2020}), \bibinfo{pages}{1--27}.
\newblock


\bibitem[Bae et~al\mbox{.}(2021)]%
        {bae2021rudra}
\bibfield{author}{\bibinfo{person}{Yechan Bae}, \bibinfo{person}{Youngsuk Kim}, \bibinfo{person}{Ammar Askar}, \bibinfo{person}{Jungwon Lim}, {and} \bibinfo{person}{Taesoo Kim}.} \bibinfo{year}{2021}\natexlab{}.
\newblock \showarticletitle{Rudra: Finding memory safety bugs in rust at the ecosystem scale}. In \bibinfo{booktitle}{\emph{Proceedings of the ACM SIGOPS 28th Symposium on Operating Systems Principles}}. \bibinfo{pages}{84--99}.
\newblock


\bibitem[Boos et~al\mbox{.}(2020)]%
        {boos2020theseus}
\bibfield{author}{\bibinfo{person}{Kevin Boos}, \bibinfo{person}{Namitha Liyanage}, \bibinfo{person}{Ramla Ijaz}, {and} \bibinfo{person}{Lin Zhong}.} \bibinfo{year}{2020}\natexlab{}.
\newblock \showarticletitle{Theseus: an experiment in operating system structure and state management}. In \bibinfo{booktitle}{\emph{14th USENIX Symposium on Operating Systems Design and Implementation (OSDI 20)}}. \bibinfo{pages}{1--19}.
\newblock


\bibitem[Burtsev et~al\mbox{.}(2021)]%
        {burtsev2021isolation}
\bibfield{author}{\bibinfo{person}{Anton Burtsev}, \bibinfo{person}{Dan Appel}, \bibinfo{person}{David Detweiler}, \bibinfo{person}{Tianjiao Huang}, \bibinfo{person}{Zhaofeng Li}, \bibinfo{person}{Vikram Narayanan}, {and} \bibinfo{person}{Gerd Zellweger}.} \bibinfo{year}{2021}\natexlab{}.
\newblock \showarticletitle{Isolation in Rust: What is Missing?}. In \bibinfo{booktitle}{\emph{Proceedings of the 11th Workshop on Programming Languages and Operating Systems}}. \bibinfo{pages}{76--83}.
\newblock


\bibitem[Contributors(2024a)]%
        {asterinas_github}
\bibfield{author}{\bibinfo{person}{Asterinas Contributors}.} \bibinfo{year}{2024}\natexlab{a}.
\newblock \bibinfo{title}{Asterinas - A {Rust}-based operating system}.
\newblock
\newblock
\urldef\tempurl%
\url{https://github.com/asterinas/asterinas}
\showURL{%
\tempurl}
\newblock
\shownote{[Accessed: 2024-05-20]}.


\bibitem[Contributors(2024b)]%
        {rCore_github}
\bibfield{author}{\bibinfo{person}{{rCore} Contributors}.} \bibinfo{year}{2024}\natexlab{b}.
\newblock \bibinfo{title}{{rCore} operating system}.
\newblock
\newblock
\urldef\tempurl%
\url{https://github.com/rcore-os/rCore}
\showURL{%
\tempurl}
\newblock
\shownote{[Accessed: 2024-05-20]}.


\bibitem[Cui et~al\mbox{.}(2023)]%
        {cui2023safedrop}
\bibfield{author}{\bibinfo{person}{Mohan Cui}, \bibinfo{person}{Chengjun Chen}, \bibinfo{person}{Hui Xu}, {and} \bibinfo{person}{Yangfan Zhou}.} \bibinfo{year}{2023}\natexlab{}.
\newblock \showarticletitle{SafeDrop: Detecting memory deallocation bugs of rust programs via static data-flow analysis}.
\newblock \bibinfo{journal}{\emph{ACM Transactions on Software Engineering and Methodology}} \bibinfo{volume}{32}, \bibinfo{number}{4} (\bibinfo{year}{2023}), \bibinfo{pages}{1--21}.
\newblock


\bibitem[Cui et~al\mbox{.}(2024)]%
        {cui2024unsafe}
\bibfield{author}{\bibinfo{person}{Mohan Cui}, \bibinfo{person}{Shuran Sun}, \bibinfo{person}{Hui Xu}, {and} \bibinfo{person}{Yangfan Zhou}.} \bibinfo{year}{2024}\natexlab{}.
\newblock \showarticletitle{Is unsafe an Achilles' Heel? A Comprehensive Study of Safety Requirements in Unsafe Rust Programming}. In \bibinfo{booktitle}{\emph{Proceedings of the IEEE/ACM 46th International Conference on Software Engineering}}. \bibinfo{pages}{1--13}.
\newblock


\bibitem[Evans et~al\mbox{.}(2020)]%
        {evans2020rust}
\bibfield{author}{\bibinfo{person}{Ana~Nora Evans}, \bibinfo{person}{Bradford Campbell}, {and} \bibinfo{person}{Mary~Lou Soffa}.} \bibinfo{year}{2020}\natexlab{}.
\newblock \showarticletitle{Is \text{Rust} used safely by software developers?}. In \bibinfo{booktitle}{\emph{Proceedings of the ACM/IEEE 42nd International Conference on Software Engineering}}. \bibinfo{pages}{246--257}.
\newblock


\bibitem[Fulton et~al\mbox{.}(2021)]%
        {fulton2021benefits}
\bibfield{author}{\bibinfo{person}{Kelsey~R Fulton}, \bibinfo{person}{Anna Chan}, \bibinfo{person}{Daniel Votipka}, \bibinfo{person}{Michael Hicks}, {and} \bibinfo{person}{Michelle~L Mazurek}.} \bibinfo{year}{2021}\natexlab{}.
\newblock \showarticletitle{Benefits and drawbacks of adopting a secure programming language: Rust as a case study}. In \bibinfo{booktitle}{\emph{Seventeenth Symposium on Usable Privacy and Security (SOUPS 2021)}}. \bibinfo{pages}{597--616}.
\newblock


\bibitem[Group(2024a)]%
        {safeandunsafe}
\bibfield{author}{\bibinfo{person}{Rust Group}.} \bibinfo{year}{2024}\natexlab{a}.
\newblock \bibinfo{title}{How Safe and Unsafe Interact}.
\newblock
\newblock
\urldef\tempurl%
\url{https://doc.rust-lang.org/nomicon/safe-unsafe-meaning.html}
\showURL{%
\tempurl}
\newblock
\shownote{[Accessed: 2024-06-07]}.


\bibitem[Group(2024b)]%
        {unsaferust}
\bibfield{author}{\bibinfo{person}{Rust Group}.} \bibinfo{year}{2024}\natexlab{b}.
\newblock \bibinfo{title}{Unsafe Rust}.
\newblock
\newblock
\urldef\tempurl%
\url{https://doc.rust-lang.org/book/ch19-01-unsafe-rust.html}
\showURL{%
\tempurl}
\newblock
\shownote{[Accessed: 2024-06-07]}.


\bibitem[Jung et~al\mbox{.}(2019)]%
        {jung2019stacked}
\bibfield{author}{\bibinfo{person}{Ralf Jung}, \bibinfo{person}{Hoang-Hai Dang}, \bibinfo{person}{Jeehoon Kang}, {and} \bibinfo{person}{Derek Dreyer}.} \bibinfo{year}{2019}\natexlab{}.
\newblock \showarticletitle{Stacked borrows: an aliasing model for Rust}.
\newblock \bibinfo{journal}{\emph{Proceedings of the ACM on Programming Languages}} \bibinfo{volume}{4}, \bibinfo{number}{POPL} (\bibinfo{year}{2019}), \bibinfo{pages}{1--32}.
\newblock


\bibitem[Jung et~al\mbox{.}(2017)]%
        {jung2017rustbelt}
\bibfield{author}{\bibinfo{person}{Ralf Jung}, \bibinfo{person}{Jacques-Henri Jourdan}, \bibinfo{person}{Robbert Krebbers}, {and} \bibinfo{person}{Derek Dreyer}.} \bibinfo{year}{2017}\natexlab{}.
\newblock \showarticletitle{RustBelt: Securing the foundations of the Rust programming language}.
\newblock \bibinfo{journal}{\emph{Proceedings of the ACM on Programming Languages}} \bibinfo{volume}{2}, \bibinfo{number}{POPL} (\bibinfo{year}{2017}), \bibinfo{pages}{1--34}.
\newblock


\bibitem[Li et~al\mbox{.}(2021)]%
        {li2021mirchecker}
\bibfield{author}{\bibinfo{person}{Zhuohua Li}, \bibinfo{person}{Jincheng Wang}, \bibinfo{person}{Mingshen Sun}, {and} \bibinfo{person}{John~CS Lui}.} \bibinfo{year}{2021}\natexlab{}.
\newblock \showarticletitle{MirChecker: detecting bugs in Rust programs via static analysis}. In \bibinfo{booktitle}{\emph{Proceedings of the 2021 ACM SIGSAC conference on computer and communications security}}. \bibinfo{pages}{2183--2196}.
\newblock


\bibitem[Liu et~al\mbox{.}(2020)]%
        {liu2020securing}
\bibfield{author}{\bibinfo{person}{Peiming Liu}, \bibinfo{person}{Gang Zhao}, {and} \bibinfo{person}{Jeff Huang}.} \bibinfo{year}{2020}\natexlab{}.
\newblock \showarticletitle{Securing unsafe rust programs with XRust}. In \bibinfo{booktitle}{\emph{Proceedings of the ACM/IEEE 42nd International Conference on Software Engineering}}. \bibinfo{pages}{234--245}.
\newblock


\bibitem[Matsushita et~al\mbox{.}(2022)]%
        {matsushita2022rusthornbelt}
\bibfield{author}{\bibinfo{person}{Yusuke Matsushita}, \bibinfo{person}{Xavier Denis}, \bibinfo{person}{Jacques-Henri Jourdan}, {and} \bibinfo{person}{Derek Dreyer}.} \bibinfo{year}{2022}\natexlab{}.
\newblock \showarticletitle{RustHornBelt: a semantic foundation for functional verification of Rust programs with unsafe code}. In \bibinfo{booktitle}{\emph{Proceedings of the 43rd ACM SIGPLAN International Conference on Programming Language Design and Implementation}}. \bibinfo{pages}{841--856}.
\newblock


\bibitem[McCormack et~al\mbox{.}(2024)]%
        {mccormack2024against}
\bibfield{author}{\bibinfo{person}{Ian McCormack}, \bibinfo{person}{Tomas Dougan}, \bibinfo{person}{Sam Estep}, \bibinfo{person}{Hanan Hibshi}, \bibinfo{person}{Jonathan Aldrich}, {and} \bibinfo{person}{Joshua Sunshine}.} \bibinfo{year}{2024}\natexlab{}.
\newblock \showarticletitle{" Against the Void": An Interview and Survey Study on How Rust Developers Use Unsafe Code}.
\newblock \bibinfo{journal}{\emph{arXiv preprint arXiv:2404.02230}} (\bibinfo{year}{2024}).
\newblock


\bibitem[Narayanan et~al\mbox{.}(2020)]%
        {narayanan2020redleaf}
\bibfield{author}{\bibinfo{person}{Vikram Narayanan}, \bibinfo{person}{Tianjiao Huang}, \bibinfo{person}{David Detweiler}, \bibinfo{person}{Dan Appel}, \bibinfo{person}{Zhaofeng Li}, \bibinfo{person}{Gerd Zellweger}, {and} \bibinfo{person}{Anton Burtsev}.} \bibinfo{year}{2020}\natexlab{}.
\newblock \showarticletitle{$\{$RedLeaf$\}$: isolation and communication in a safe operating system}. In \bibinfo{booktitle}{\emph{14th USENIX Symposium on Operating Systems Design and Implementation (OSDI 20)}}. \bibinfo{pages}{21--39}.
\newblock


\bibitem[Qin et~al\mbox{.}(2024)]%
        {qin2024understanding}
\bibfield{author}{\bibinfo{person}{Boqin Qin}, \bibinfo{person}{Yilun Chen}, \bibinfo{person}{Haopeng Liu}, \bibinfo{person}{Hua Zhang}, \bibinfo{person}{Qiaoyan Wen}, \bibinfo{person}{Linhai Song}, {and} \bibinfo{person}{Yiying Zhang}.} \bibinfo{year}{2024}\natexlab{}.
\newblock \showarticletitle{Understanding and Detecting Real-World Safety Issues in Rust}.
\newblock \bibinfo{journal}{\emph{IEEE Transactions on Software Engineering}} (\bibinfo{year}{2024}).
\newblock


\bibitem[Qin et~al\mbox{.}(2020)]%
        {qin2020understanding}
\bibfield{author}{\bibinfo{person}{Boqin Qin}, \bibinfo{person}{Yilun Chen}, \bibinfo{person}{Zeming Yu}, \bibinfo{person}{Linhai Song}, {and} \bibinfo{person}{Yiying Zhang}.} \bibinfo{year}{2020}\natexlab{}.
\newblock \showarticletitle{Understanding memory and thread safety practices and issues in real-world Rust programs}. In \bibinfo{booktitle}{\emph{Proceedings of the 41st ACM SIGPLAN Conference on Programming Language Design and Implementation}}. \bibinfo{pages}{763--779}.
\newblock


\bibitem[Rivera et~al\mbox{.}(2021)]%
        {rivera2021keeping}
\bibfield{author}{\bibinfo{person}{Elijah Rivera}, \bibinfo{person}{Samuel Mergendahl}, \bibinfo{person}{Howard Shrobe}, \bibinfo{person}{Hamed Okhravi}, {and} \bibinfo{person}{Nathan Burow}.} \bibinfo{year}{2021}\natexlab{}.
\newblock \showarticletitle{Keeping safe rust safe with galeed}. In \bibinfo{booktitle}{\emph{Proceedings of the 37th Annual Computer Security Applications Conference}}. \bibinfo{pages}{824--836}.
\newblock


\bibitem[Xu et~al\mbox{.}(2021)]%
        {xu2021memory}
\bibfield{author}{\bibinfo{person}{Hui Xu}, \bibinfo{person}{Zhuangbin Chen}, \bibinfo{person}{Mingshen Sun}, \bibinfo{person}{Yangfan Zhou}, {and} \bibinfo{person}{Michael~R Lyu}.} \bibinfo{year}{2021}\natexlab{}.
\newblock \showarticletitle{Memory-safety challenge considered solved? An in-depth study with all Rust CVEs}.
\newblock \bibinfo{journal}{\emph{ACM Transactions on Software Engineering and Methodology (TOSEM)}} \bibinfo{volume}{31}, \bibinfo{number}{1} (\bibinfo{year}{2021}), \bibinfo{pages}{1--25}.
\newblock


\bibitem[Yu et~al\mbox{.}(2019)]%
        {yu2019fearless}
\bibfield{author}{\bibinfo{person}{Zeming Yu}, \bibinfo{person}{Linhai Song}, {and} \bibinfo{person}{Yiying Zhang}.} \bibinfo{year}{2019}\natexlab{}.
\newblock \showarticletitle{Fearless concurrency? understanding concurrent programming safety in real-world rust software}.
\newblock \bibinfo{journal}{\emph{arXiv preprint arXiv:1902.01906}} (\bibinfo{year}{2019}).
\newblock


\end{thebibliography}
\end{document}